%% file: ms.tex
\documentclass{emulateapj}

\usepackage{subfigure}
\usepackage{ifthen}

\newboolean{preprint}
\setboolean{preprint}{false}

\input mymac.tex

\slugcomment{Accepted for publication in ApJ.}
\shorttitle{Conditions in Low-Ionization Starburst Outflows}
\shortauthors{Martin and Bouch\'e}

\begin{document}

\title{Physical Conditions in the Low-Ionization Component of
Starburst Outflows: The Shape of Near-Ultraviolet and Optical
Absorption-Line Troughs in Keck Spectra of ULIRGs}

\author{Crystal L. Martin\altaffilmark{1}}
\affil{University of California, Santa Barbara}
\affil{Department of Physics}
\affil{Santa Barbara, CA, 93106}
\email{cmartin@physics.ucsb.edu}

\and
\author{Nicolas Bouch\'e}
\affil{Max-Planck\_Institut F\"ur Extraterrestrische Physik (MPE)}
\affil{Giessenbachstr. 1, 85748}
\affil{Garching, Germany}


\altaffiltext{1}{Packard Fellow}


\begin{abstract}
We analyze the physical conditions in the low-ionization component of starburst
outflows  (in contrast to the high-ionization wind fluid observed in X-rays), 
based on new Keck/LRIS spectroscopy of partially resolved absorption troughs in 
near-ultraviolet and optical spectra of Ultraluminous Infrared Galaxies. The large 
velocity width and blueshift present in seven, atomic transitions indicate a
macroscopic velocity gradient within the outflowing gas. The separation of 
the \mgII 2796, 2803 (and \feII 2587, 2600) doublet lines in these data constrains the 
gas kinematics better than previous studies of the heavily blended \nad 5892, 98 doublet. 
The identical shape of the \mgII 2796 absorption troughs to that of the normally 
weaker transition at 2803\AA\ (after accounting for emission filling) requires both 
transitions be optically thick at all outflow velocities. The fraction of the 
galactic continuum covered by the outflow at each velocity therefore dictates the
shape of these absorption troughs. We suggest that the velocity offset of the deepest 
part of the troughs, where the covering factor of low-ionization gas is near unity, 
reflects the speed of a shell of swept-up, interstellar gas at the time of blowout. 
In a spherical outflow, we show that the fragments of this shell, or any clouds that 
expand adiabatically in rough pressure equilibrium with the hot wind, expand slowly 
relative to the geometrical dilution; and the covering fraction of low-ionization 
gas decreases with increasing radius. Our measurement of a covering factor that 
decreases with increasing velocity can therefore be interpreted as evidence that the 
low-ionization outflow is accelerating, i.e. absorption at higher velocity comes from gas 
at larger radii. We also present measurements of $C_f(v)$ in 4 species, place an
upper limit of $n_e \sles\ 3000$\cm3 on the density of the outflowing gas, and
discuss lower limits on the mass outflow rate.

\end{abstract}


\section{INTRODUCTION}

Galactic winds are a key ingredient in cosmological models of
galaxy evolution. They transport the nucleosynthetic products
of stars into galaxy halos and the intergalactic medium (IGM),
thereby shaping enrichment and effective yields. The amount
of material ejected from low mass galaxies can significantly 
reduce the gas mass available for star formation, resulting
in lower global efficiencies for star formation that flatten
the faint end of the galaxy luminosity function, relative 
to that of the halo mass function (Somerville \& Primack 1999).  Some
heating mechanism is apparently required in very massive
halos in order to make the number of galaxies cutoff more
steeply with luminosity than the halo mass distribution does
with increasing halo mass. Simulations generally invoke some
type of feedback from accreting supermassive black holes, 
although only the energy scale required distinguishes this
feedback from supernova-driven, galactic winds. With mass-loading
factors that scale inversely with galaxy mass,
feedback prescriptions, starburst winds account for the metal 
enrichment of the intergalactic medium (IGM;e.g., 
Oppenheimer \& Dav\'e 2006).

The scaling relations used for feedback in cosmological 
simulations are largely motivated by empirical results. 
Nearby starburst galaxies present hot, extraplanar gas,
heated by supernovae (Dahlem \et 1998; Martin 1999; Strickland 
\et 2002; Martin, Kobulnicky, \& Heckman 2002).
Outflow speeds for low-ionization gas have been measured for much
larger samples of nearby galaxies (Heckman \et 2000; Rupke, Veilleux, 
\& Sanders 2002; Martin 2005; Rupke, Veilleux, \& Sanders 2005b; 
Sato \et 2009) and composite spectra of distant galaxies 
(Shapley \et 2003; Weiner \et 2009).  In the standard dynamical
model for the acceleration of this gas, the kinematics of the 
low-ionization gas reflect properties of a hotter, energetically 
dominant wind. The hot wind is not a necessary component of the outflow 
when radiation pressure (on dust grains) accelerates the outflow. 
Outflow components at vastly different temperatures can only be observed 
from the X-ray to the infrared for relatively nearby systems. 
At redshifts $z \sgreat 0.15$, observations of winds typically detect
low-ionization absorption lines and, less commonly, cover  strong
transitions of \oVI, NV, SiIV, and CIV. 
{\it Determining what physical information 
can be reliably extracted from the absorption lines is of paramount importance for 
understanding any evolution in galactic wind properties over cosmic time.}

For nearby starbursts, the \nad 5892, 5898 lines are the most commonly detected
lines from outflows owing to the aperture advantage of ground-based telescopes over 
satellites. The majority of luminous starburst galaxies present a strong, blueshifted 
\nad\ absorption trough  (Heckman \et 2000; Rupke \et 2002; Martin 2005, 2006; 
Rupke \et 2005b; Sato \et 2009); but the detected fraction in \nad\ drops to about
half for the nearby dwarf, starburst galaxies (Schwartz \& Martin 2004).  
Although the stronger line, \nad 5892, is usually saturated based on its
strength relative to the weaker line at 5898\AA, the absorption troughs are rarely 
black. These previous studies have concluded that the \nad absorbers do not 
completely cover the optical continuum source. The blended absorption troughs
have been fitted with the minimum number of velocity components, typically from 1 to 3,
required to describe the velocity asymmetry (e.g. Martin 2005, 2006; Rupke \et 2002, 2005a);
and little has been written about the variation of physical parameters with velocity
in the outflow. In contrast, the very broad absorption troughs identified as outflows in active
galactic nuclei (AGNs)
spectra require a  velocity-dependent covering factor (Arav \et 1999a, 2001a; de Kool \et
2002; Gabel \et 2003; Scott \et 2004). Resolving absorption troughs in starburst spectra,
and comparing different transitions, can lead to new insight about the nature of
the low-ionization component of outflows.

In this paper, we present Keck/LRIS spectroscopy of redshift $z \sim 0.25$ 
galaxies that partially resolves the \nad 5892, 5897, \mgI 2853, \mgII 2796,
2803, and \feII 2587, 2600 absorption troughs. We choose Ultraluminous Infrared Galaxies (ULIRGs)
 due to their high 
\nad outflow fraction and selected systems at redshifts that provided \nad and near-ultraviolet
spectral coverage. The targets are among the most luminous starbursts in the local 
universe and are classified as ULIRGs, $\log L_{IR} > 12$.
Their activity triggers by gas inflow induced by a recent, or on-going, merger 
(Borne \et 2000).  The wider separation of the \mgII doublet lines, 768\kms, relative to the 
\nad doublet, allows direct comparison of the  $\lambda 2796$ and $\lambda 2803$ troughs, 
hereafter the {\em blue} and {\em red} troughs respectively. At any location in the outflow, the 
blue transition has twice the optical depth of the red one; and the covering fraction will 
necessarily be identical. In contrast to the $\alpha$-element enrichment measured in the
hot wind (Martin, Kobulnicky, \& Heckman 2002), the relative abundance of Fe to Mg could be as 
high as solar in the low-ionization gas, which may be primarily entrained interstellar gas.
The optical depth in the \feII line would then be just 1.8 times lower than that in \mgII 2803.
However, the much weaker \feII 2587 line, $\tau_0(2600) = 3.5 \tau_0(2587) $, might be optically
thin when the other transitions are saturated. Whether \feII 2587 or \nad 5898 proves to
have lower optical depth, and therefore provides the better measurement of ionic column
density, depends on the relative ionization corrections for \feII and \nad.

Our presentation is organized as follows. Section~\ref{sec:data} describes the new
observations, derives emission-line redshifts, and estimates the sensitivity to 
outflowing gas. On the first reading, we recommend skipping directly to \S~\ref{sec:results}, 
where the absorption trough measurements are first presented 
in a model-independent fashion. Section~\ref{sec:parametric} then uses the independent 
measurements of the trough intensity $I_B(v)$ and $I_R(v)$ to solve directly for
optical depth $\tau_R(v)$ and covering fraction $C_f(v)$ as a function of outflow velocity.
Since this approach cannot be applied to the majority of the absorption troughs, we
introduce a $\chi_{\nu}^2$-fitting method in \S~\ref{sec:chi2}. It differs from previous
analyses in terms of the set of physically motivated priors adopted. The advantages
are quantitative measurements of the velocity dependence of the covering factor in
four species and direct comparison of the limits on gas column density from different
transitions. In this way, we establish the relation of the outflow component probed
by near-UV resonance lines to that seen in \nad. The connection enables comparison of wind 
properties over a large redshift range. 
Our discussion in Section~\ref{sec:discussion} focuses on interpreting the velocity-dependence
of the covering fraction. The main results and their implications are summarized in
Section~\ref{sec:summary}.

Oscillator strengths and vacuum wavelengths are taken from Morton (1991, 2003) for the optical 
and near-ultraviolet transitions. We adopt a cosmological model with
$H_0 = 70$\kms ~Mpc$^{-1}$, $\Omega_0 = 0.3$, and $\Omega_{\Lambda} = 0.7$
throughout the paper.

\section{KECK OBSERVATIONS OF OUTFLOWS IN THE NEAR-ULTRAVIOLET AND OPTICAL} \label{sec:data}

We obtained (rest-frame)  near-UV and optical spectroscopy of redshift $\sim 0.25$
starburst galaxies with LRIS (Oke \et 1995; McCarthy \et 1998) on Keck~I. 
The blue channel of LRIS is one of the few 
optical spectrographs that is efficient down to the atmospheric cut-off 
in the blue, allowing observations of MgII at $z \sgreat\ 0.15$
(for targets directly overhead).  The dual-beam design offers simultaneous 
coverage of NaI absorption and several emission lines in the red channel. 
Due to the low density of bright galaxies at these redshifts, the spectra 
must be obtained one at time. 

Galaxies were chosen from the {\sc IRAS} 1~Jy survey of ULIRGs
(Kim \& Sanders 1998), which contains 118 ULIRGs with $F_{60} > 1$~Jy,
based on apparent magnitude and airmass at the time of our runs.
Ultraluminous Infrared Galaxies are extremely rare at the present 
epoch but offer the closest local analog of the IR-luminous galaxy
population that dominates the cosmic star formation rate density 
at redshifts greater than 0.7 (Le Floc'h \et 2005). The high dust content absorbs much of 
the stellar luminosity and emits this energy as thermal radiation in 
the infrared.

Table~\ref{tab:sample} lists some properties of the galaxies whose spectra
are presented in this paper. The infrared luminosities of the galaxies observed 
range from $\log L_{IR} = 12.31 - 12.81$ with mean  $\log L_{IR} = 12.58$, 
which is more luminous than the mean of the Martin (2005, 2006) sample.
The AGN fraction is higher in brighter ULIRGs (Veilleux \et 1999); and
two of the five ULIRGs in the sample are spectroscopically classified as AGN 
(2 LINERs and 2 Sey~2's). The more recent Kewley \et (2006) division of these
optical-emission-line ratios into excitation types classifies 
FSC~0039-13 as a composite starburst-AGN, FSC~1009+47 as a Seyfert~2, FSC~1407+05
at the Seyfert/LINER boundary, FSC~1630+15 at the LINER/Composite boundary, and
FSC~2349+24 as a Seyfert~2. Among our five targets, FSC~2349+24 clearly presents
the strongest AGN signatures, but AGN contribute less than half of the bolometric
luminosity in the other four ULIRGs (Veilleux \et 2009a). Among our five targets, FSC~2349+24 clearly presents
the strongest AGN signatures.
Rupke \et (2005d) found little difference in NaI wind 
kinematics between HII, LINER, and Sey 2 ULIRGs; only the Sey 1's showed faster 
outflows. Hence, we expect these outflows are driven primarily by the starburst 
rather than the AGN. The far-infrared colors are all cooler than the standard
demarcation, $F(25\mu m)/ F(60\mu m) > 0.2$, used to identify warm ULIRGs 
(Kim \et 1998), which also suggests the total luminosity is starburst dominated.
The total infrared luminosities correspond to upper 
limits on the star formation rates from $SFR \sim 207 - 653$\msunyr\ 
or $\sim 352-1110$\msunyr, respectively, for the Chabrier or Salpeter 
initial mass functions.

\subsection{Data Acquisition and Reduction}

Longslit spectra were obtained with LRIS (Oke \et 1995) and LRISb 
(McCarthy \et 1998) on 2004 January 26, 2004 March 16-17, 
2007 October 6, and 2007 November 1. Clouds and high humidity 
limited the exposure times on all of these nights. We present the 
five highest quality spectra obtained from the combined data.

The position angle of the longslit was selected to cover both nuclei
of FSC~1009+47  and FSC~2349+24. The atmospheric dispersion compensator,
which was installed on LRIS in spring 2007, was used for the 2007
observations.  Care was taken with the 2004 observations to observe
the targets when the slit PA was near the parallactic angle. 

The dual beam spectrograph was configured with the D560 dichroic,
a 1200-l grism blazed at 3400\AA\ in the blue arm (LRISb) and
a 1200-l grating blazed at 7500\AA\ in the red arm. The grating
tilt for each target was tuned to cover the \Ha + [NII] emission
lines as well as the NaI absorption and HeI 5876 emission line.
The blue spectra cover the \mgII 2796, 2803 doublet, the \mgI 2853 line, 
and the \feII 2587, 2600 doublet . The slit width was chosen based on the atmospheric 
seeing.  The resolution (for a source
filling the slit) ranged from 110\kms on the 2004 March run to 160\kms 
on the 2007 November run and is listed in Table~\ref{tab:data}.
The size (FWHM) of the galaxies along the slit exceeded the slit
width. 

Fixed-pattern noise was removed using software scripts that
called {\sc IRAF} tasks.\footnote{IRAF is distributed by the National 
        Optical Astronomy Observatory, which is operated by the 
	Association of Universities for Research in Astronomy (AURA) 
	under cooperative agreement with the National Science Foundation.} 
The blue spectra were flatfielded with twilight sky frames, normalized by 
the sky spectrum. The red spectra were flatfielded using an internal 
(spectrosopic) exposure of a quartz lamp. 
A dispersion solution was fitted to the vacuum wavelengths of
the arc lamp lines as a function of detector coordinate. The root mean 
square error in the dispersion solution for the blue (red) spectra was 
0.05 (0.07) \AA. Application of a small, additive shift, up to a couple 
tenths of an angstrom, registered the wavelengths of night sky emission lines 
with their values in a telluric-line spectrum (Hanuschik 2003) smoothed to our 
spectral resolution and transformed to vacuum wavelengths (using the Edlen formula). 
We attribute these corrections to shifts in the dispersion solution with airmass 
and rotator angle. The correction to the Local Standard of Rest (LSR)  was 
computed for each observation using the {\sc IRAF} task {\sc RVCORRECT}. All 
the offsets were less than 30\kms. Since we are only concerned with relative 
velocites, the corrections to LSR were not applied to the data.

We rectified the two-dimensional spectral images, using the dispersion solution 
and traces of a standard star stepped along the longslit, and then extracted
an integrated galaxy spectrum for each target. These spectra 
have $SNR \sim 5 - 10$ per pixel as shown in Table~\ref{tab:data}. Additional,
lower quality spectra were extracted for the distinct continuum sources
within FSC~1009+47, FSC~1407+05, and FSC~2349+24. After substraction of the 
median sky intensity at each wavelength, significant residuals from
strong night-sky emission lines remained 
in the red spectra near the \nad line in FSC~1009+47, 1407+05, and 1630+15.
Variance spectra were extracted for each target prior to sky 
subtraction and flux calibration. The variance vectors were later
scaled by a multiplicative factor that made the uncertainties in
the intensity consistent with the measured standard deviation in
the extracted, target spectra.

Observations of multiple, standard stars determined the relative sensitivity 
with wavelength. The data were flux calibrated using this sensitivity
function and then normalized by a fitted continuum. The error in continuum
placement is neligible around \nad, \mgI, and \mgII; but the blending of stellar 
absorption lines washes out the true continuum level at shorter wavelengths.  
To identify bandpasses near \feII 2587, 2600 that likely reach the true 
continuum level, we compared high-resolution, synthesized spectra models of
stellar populations to copies smoothed to 100\kms resolution. We fitted a first or second
order cubic-spline through these bandpasses as well as the broad bandpasses near 
the \mgI and \mgII lines. The resulting error in the continuum level near \feII 
depends on the star formation history. The severe blanketing in older bursts requires
actual fitting of reddened, model spectra. This level of sophistication was unneccessary,
however, because these near-UV spectra are bluer than the $t > 100$~Myr burst models; and
we confidently rule out old, burst models for the continuum. The population synthesis 
models indicate either continuous star formation or a burst within the past $100$~Myr.
Repeated fitting trials indicate the uncertainty in the continuum level around \feII 
was 1 to  5\%. 

For a typical line width of 470\kms FWHM, we detect a rest-frame 
equivalent width $W_r(5\sigma) \approx 1.52 \AA (5/SNR_{pix}) (\Delta v / 
470~{\rm km~s}^{-1})^{0.5} $ in the red spectra at the $5\sigma$ 
significance level. A typical $5\sigma$ sensitivity limit for the 
blue spectra is  $W_r(5\sigma) \approx 0.92 \AA (5/SNR_{pix}) (\Delta v 
/ 470 ~{\rm km~s}^{-1})^{0.5}$.

\begin{figure}
\hspace{2cm}
\hbox{See Figure 1 on p.22}
\caption{Normalized intensity vs. velocity, where the velocity is relative 
to the systemic velocity determined from emission lines.
\label{fig:spec}
}
\end{figure}

\subsection{Sensitivity of Spectra to Physical Properties}

We measured redshifts from recombination lines of {\rm H} and 
{\rm He} and forbidden lines of singly-ionized {\rm N} and {\rm S}.
These lines all fall in our red spectra. The  dispersion solution ties
both the blue and the red spectra to the {\em vacuum wavelengths} 
of night sky emission lines. These redshifts agree with the previous 
measurements of Kim \& Sanders (1998). Our independent check is important 
because that work used the \nad\ absorption line in the redshift estimate;
and  we will show that the \nad kinematics differ significantly from that
of the recombination lines.

The sensitivity of the seven metal lines to outflowing gas depends
on the relative abundances of the elements, the oscillator strengths of 
the transitions, the dust depletion, and the ionization corrections.  In
the limit of comparable ionization corrections for all species, the optical 
depth in the \mgI  2853 line would be the largest for solar-abundance 
ratios. The \mgII lines have lower oscillator strengths, and the cosmic abudances 
of Fe and Na are lower. Our typical detection limit for \mgI corresponds to a minimum 
hydrogen column density,
\begin{eqnarray}
 N(HI) & =&  1.62 \times 10^{17} {\rm ~cm}^{-2} (N(Mg)/N(MgI)) ~ \times \nonumber \\
&& (5/SNR)~\frac{1.25}{1+z}(\Delta v / 470 {\rm ~km~s}^{-1} )^{0.5}.
\end{eqnarray}
Under the same conditions, the \nad 5898 line optical depth would be slightly lower 
than the weakest \feII line, but whether either transition is optically-thin depends 
on the relative ionization correction, where
\begin{eqnarray}
\frac{\tau_0(\feII 2587)}{\tau_0(\nad 5898)} = 1.41 \frac{\chi(\feII)}{\chi(\nad)}.
\end{eqnarray}
The \nad 5898 line can, in principle, provide the best measurements of an ionic column
density in the outflow, but interpretation is complicated by the blending with the
\nad 5892 transition, the ionization correction, and depletion of Na by dust grains.
The \feII transitions with the lowest oscillator strengths lie blueward of the 
atmospheric cut-off for nearby galaxies, and their measurement will likely lead to 
the best estimates of mass column density.

\section{MEASURED PROPERTIES OF LOW-IONIZATION OUTFLOWS} \label{sec:results}

\fig~\ref{fig:spec} shows the absorption troughs on a velocity scale.
The high spectral resolution, relative to the broad troughs, allows us to compare 
the shape of the absorption troughs among seven transitions. The centroids of the 
\Ha and [NII] emission lines lie at the systemic velocity by construction. 
The spectra of all five galaxies present broad absorption troughs in the \nad\ and 
\mgII\ doublets.  The spectrum of FSC~0039-13 has the best SNR near the \mgI\ 
2853, \feII\ 2600, and \feII\ 2587 lines; but these transitions are detected 
in all the spectra except for FSC~2349+24. The \feII\ 2587, 2600 lines lie shortward of 
$\lambda 3200$ for FSC~2349+24, and the SNR is severely compromised by atmospheric attenuation. 
In Figure~\ref{fig:spec}, the two Seyfert~2 galaxies in the sample, FSC~1407+05 and FSC~2349+24, 
present \mgII in emission. Some He~I emission is present near the \nad doublet in all the 
spectra and is most prominent in FSC~0039-13.

Spectra of A, F, and G stars present low-ionization, metal lines; but any 
photospheric contribution to the total equivalent width in our spectra is small.
The velocity offset of the absorption troughs require a non-stellar absorption component;
and  the \mgII troughs for four of the galaxies show little or no absorption at the 
systemic velocity. Only the FSC~0039-13 spectrum presents absorption at $v=0$.
We attribute it to the interstellar medium in the host galaxy, likely a result
of high inclination relative to our sightline.

\begin{figure}[h]
\centering
\includegraphics[scale=0.45,angle=0,clip=true]{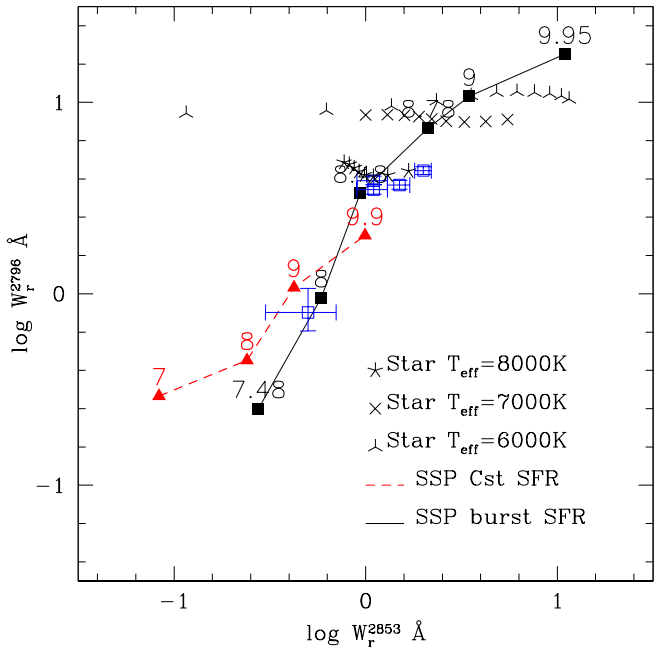}
\caption{\mgI and \mgII absorption equivalent widths for 5 ULIRGs compared to simple stellar 
population models (\S~\ref{sec:overview}) and individual stellar templates.
The equivalent widths increase with increasing age, from $\log \tau (yr) = 7$ to 
10, for both burst (solid squares) and continuous (solid triangles)
star formation histories. In the individual stellar templates, the EW(\mgII) is set by 
the effective temperature, while the EW(MgI) varies significantly with the gravity parameter. 
In \S~\ref{sec:overview}, we argue that neither the \mgI nor \mgII absorption equivalent widths 
are stellar-dominated in any of these ULIRGs.
}
\label{fig:uvblue_ew} \end{figure}

\begin{figure}[h]
\centering
\includegraphics[scale=0.35,angle=-90,clip=true]{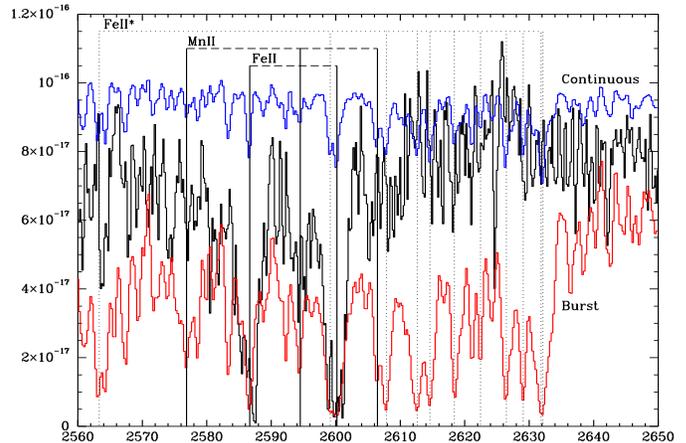}
\caption{Unreddened 
STARS2002/UVBLUE models for a 1~Gyr old stellar population of solar metallicity
vs. the FSC~0039-13 spectrum (in black) around \feII.
The data show very strong \feII lines. The weaker \mnII 2576.877, 2594.499, and 2606.462
lines are expected to have equivalent width ratios of 1.8:1.4:1.0,
respectively. Since the upper limit on the \mnII 2606 strength is low, 
the observed \feII 2600 trough cannot have a significant contribution from 
an \mnII 2594 line. The spectral quality allows only upper limits to be placed on 
the strength of the excited \feII$^*$ lines.
Excited \feII$^*$ lines are prominent in the burst model and present but weaker
than our detection limit in the continuous star formation model. Models were
normalized to the data at 3000\AA. Modest reddening of the continuous star formation 
model, e.g. $A_V = 0.5$, provides a reasonable estimate of the continuum level
in the FSC~0039-13 spectrum. Burst
models with ages greater than 100~Myr are intrinsically too red to describe the
observed continuum slope.
}
\label{fig:uvblue_feii} \end{figure}

A dominant stellar origin for the resonance absorption appears unlikely in these
ULIRGs for several reasons.
 First, \fig~\ref{fig:uvblue_ew} 
shows \mgII and \mgI equivalent widths measured from synthetic spectra, computed with
the STARS2002 stellar population synthesis code (Sternberg 1998; Thornley \et 2000;
Sternberg, Pauldrach, \& Hoffmann 2003; Davies \et 2007) and 
the UVBLUE library of high-resolution stellar spectra (Rodr\'iguez-Merino \et 2005), 
for a broad range of star formation histories.\footnote{
          All models assumed a Kroupa initial mass function
          from 1 to 120\msun, solar metallicity,
          and an exponentially declining star-formation rate.}
Only post-starburst populations, with continua
dominated by A, F, or G stars, show \mgII\ 2796 equivalent widths of more than a couple 
Angstrom. The large \mgII equivalent widths of FSC~0039-13, FSC~1407+05, and FSC~1630+15 
are inconsistent with continuous star formation; and the other two ULIRGS do not lie near 
the stellar locus. Second, the ULIRG spectra lack detectable absorption from 
excited, electronic states of \feII, lines that are prominent in the spectra
of older stellar populations in Figure~\ref{fig:uvblue_feii}. In addition, 
the synthesized spectrum of the stellar population shows \mgII lines much broader 
than those from \mgI, although smearing by the motions of stars in the galaxy could 
hide this difference. Previous ULIRG outflow studies estimated the stellar 
contamination in the \nad absorption trough from the equivalent width of the excited \mgI 
triplet at $\sim 5200$~\AA\ (not covered by our spectra) and typically found a
negligible stellar contribution  (Martin 2005). 

Extracting quantitative information about the absorbing material in the outflow 
from the absorption troughs requires a physical model. 
The spectra presented in this paper resolve components with intrinsic velocity widths 
of $100$\kms or more; and neither thermal nor turbulent motion easily explains the
extremely broad, $\sim 800$\kms, velocity width of the absorption troughs.
Macroscopic variations in velocity are required and could be associated with discrete
supershells and their fragments (Fujita \et 2009), interstellar clouds over-run
by the superbubble shock (Cooper \et 2008), eddies formed at the wind -- disk
interface (Heckman \et 2000), or the velocity gradient in a smoothly accelerating
wind (Murray \et 2005). These ideas motivate descriptions with discrete velocity components,
where components correspond to individual clouds or shell fragments. 

In Section~\ref{sec:chi2}, we formally fit all the absorption troughs with such 
velocity components.  For unblended, doublets, a parametric description of the
absorption troughs requires fewer physical priors, however; and we apply this approach in 
Section~\ref{sec:parametric} to build intuition. The key constraint throughout this modeling
is the observation that many of the transitions must be optically thick, not only at the
deepest part of the absorption trough but at high outflow speed. To illustrate the
robustness of this model-indepdendent statement, we directly compare the absorption trough 
shapes among the different transitions in Section~\ref{sec:overview}.


\subsection{Direct Comparison of Absorption Trough Shape in Seven Transitions}  \label{sec:overview}

When intervening absorption is studied in quasar spectra, or the Galactic halo studied
in absorption against stellar spectra,  the angular size of the absorbing gas clouds exceeds 
that of the continuum source. Gas between the observer and 
the light source attenuates the continuum by an amount proportional to the logarithm 
of the optical depth in any transition. For a doublet, we have
\begin{eqnarray} \label{eqn:aod}
I_B(v) = I_0 e^{-\tau_B(v)} \\
I_R(v) = I_0 e^{-\tau_R(v)}.
\end{eqnarray}
For interstellar conditions, the relative optical depth between electronic transitions 
in a single ion from the  ground state, i.e. of resonance lines, is
$ \tau_B / \tau_R = (f_B \lambda_B) / (f_R \lambda_R) $. For the \nad and  \mgII doublets,
the ratio of the oscillator strengths of the blue line, $f_B$, to the red line, $f_R$,
is 2; and the wavelengths of the transitions are very similar. 
Substituion in Eqn.~\ref{eqn:aod} indicates that the relative depth of the 
continuum-normalized absorption troughs must be $I_B(v) = I_R^2(v)$.
These relations hold provided the clouds completely cover the continuum source
over a velocity range comparable to (or greater than) the spectral resolution.
In practice, when transitions with different $f$ values saturate, both troughs appear 
black in noisy spectra.

Observations of \nad absorption troughs in ULIRGs cannot directly measure the
relative intensities $I_B(v)$ to $I_R(v)$ due to the blending of the two absorption troughs.
Previous modeling of the combined trough strongly suggests, however, that the assumption of 
complete continuum coverage is not met (Martin 2005,2006; Rupke \et 2005a,b).

\subsubsection{Description of \mgII Absorption Troughs}

The \mgII absorption troughs delineate the outflow kinematics most cleanly.
In FSC0039-13,  the majority of the trough equivalent width is blueshifted, but
the minimum intensity of the trough lies near the systemic velocity.
The intensity minima are blueshifted several hundred \kms in the other \mgII spectra
with little or no absorption at the systemic velocity. In  FSC1630+15 and FSC2349+24,  the minima in 
the absorption troughs lie at  $v \sim -400$\kms.  The FSC1407+05 spectrum presents
a broad trough from -200 to -500\kms.  The  \mgII 2796 and 2803 absorption troughs blend 
together in the FSC1009+47 spectrum, possibly due solely to the lower SNR of this spectrum.
The \mgII absorption troughs have 
residual intensities $< 10\%$ in FSC0039-12 and FSC1630+15 at minimum intensity.  The residual 
intensity at line center is low but not zero in the other three spectra. The \mgII absorption troughs are  
noticeably deeper than the \mgI or \nad absorption troughs.

We can test the condition represented by Equations~3 and 4 by
comparing the  \mgII $\lambda 2796$ and $\lambda 2803$ troughs in our spectra of 
FSC0039-13, FSC1407+05, FSC1630+15, and FSC2349+24, only FSC~1009+47 presents severe
blending of the absorption components in Figure~\ref{fig:spec}. Comparison of the \mgII 2796 and 2803 
trough shapes illustrates their nearly identical intensity at most velocities. The most significant
differences appear in the FSC2349+24 and FSC1407+05 spectra where \mgII $\lambda2796$ emission 
fills in part of the $\lambda 2803$ trough. The assumption of unity covering factor implicit to the 
apparent optical depth method clearly fails. As seen in projection on the sky against the spatially 
extended galactic continuum, the low-ionization outflow  does not uniformly cover the source.


\begin{figure}[h]
\centering
\includegraphics[scale=0.6,angle=-90,clip=true]{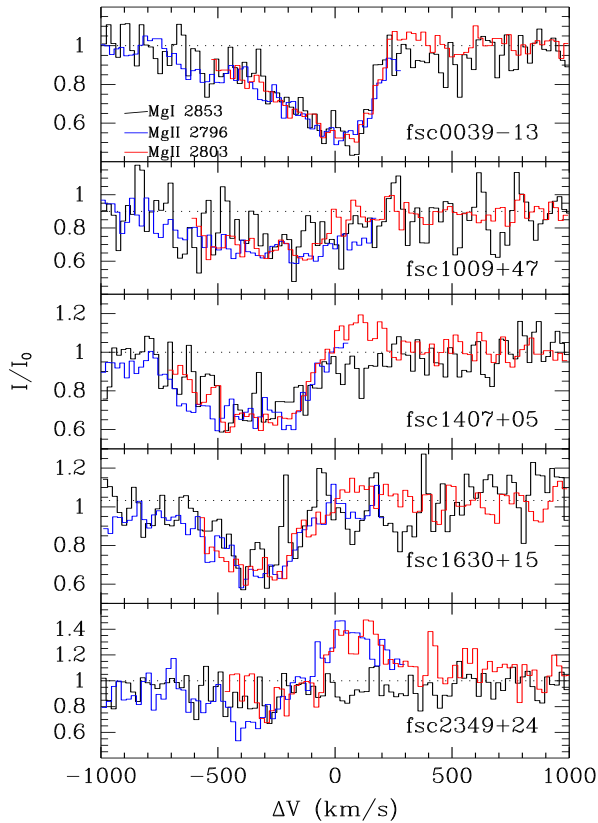}
\caption{\mgI with scaled \mgII profiles are overlaid. To facilitate comparison 
the depth  of the corresponding MgII absorption troughs have been reduced
by factors of 2.0, 2.5, 2.0, 2.0, and 1.0, respectively. The shape
of the MgII and MgI absorption troughs are indistinguishable aside from the
contribution of the \mgII emission.
}
\label{fig:mgi_mgii} \end{figure}

\subsubsection{Relative Shape of \mgI and \mgII Absorption Troughs}

In \fig~\ref{fig:mgi_mgii}, we scale the depth of each \mgII absorption 
trough to facilitate comparison to the shape of the \mgI absorption trough.
Based on
our inspection of the data, we argue that the \mgII and \mgI absorption troughs
trace gas with the same kinematics. 

\paragraph{FSC0039-13:}

We reduced the depth of the \mgII absorption trough by a factor of two by
multiplying $I_c(\lambda) - I(\lambda)$ by a factor 0.5 and adding the result to the
normalized continuum,  $I_c(\lambda)$. The absorption troughs in \mgII 2796, 2803 and \mgI 2853 are 
essentially indistinguishable after this scaling. The \mgII\ absorption 
towards FSC0039-13 is detected at velocities $\sim 100$\kms higher than 
the maximum velocities detected in \mgI. The SNR in the MgI profile 
blueward of -600\kms is inadequate to definitively detect the weak 
absorption expected based on the shape of the \mgII 2796 profile, so
the \mgI observation is consistent with the presence of neutral Mg up to the 
same outflow velocity detected in \mgII.

\paragraph{FSC1009+47:}

The scaled \mgII 2796 and 2803 absorption trough is identical to that of \mgI 
blueward of -100\kms out to -600\kms, where \mgII 2803 blends with the red part of 2796. 
The high-velocity 2803 absorption makes the 2796 trough appear to be deeper than
the 2803 trough between -100\kms\ and +150\kms. 
The 2796 trough is detected to -780\kms. The \mgI trough is
cleanly detected to -450\kms; but comparison to \mgII 2803 suggests
\mgI is plausibly present in the outflow to at least -700\kms.
After the depth of the \mgII profile is reduced by a factor of 2.5,
the \mgI profile is indistinguishable from the 2796 profile at
high velocity and the 2803 profile at low velocity.

\paragraph{FSC1407+05:}

The \mgI and \mgII 2796 lines in
FSC1407+05 are both detected to velocities reaching -750\kms.
The MgII 2803 profile presents emission from 0 to +250\kms. The 
MgII 2796 emission is diluted by the MgII 2803 absorption at the 
same wavelength. The addition of the 2796 emission to the
 \mgII 2803 absorption trough elevates the latter between 
velocities -750 and -500\kms. At intermediate velocities less
affected by blending, -500 to 0\kms, the \mgII and \mgI profiles 
lie right on top of each other after reducing the depth of the 
\mgII troughs by a factor of 2.0.

\paragraph{FSC1630+15:}

In  FSC1630+15, the shape of the \mgII\ and \mgI\ profiles are the same
from 0 to -620\kms after scaling \mgII by a factor of 2.0. Maximum 
absorption is offset to -400\kms. The \mgII 2796 absorption trough is 
detected to -825\kms. The  SNR  in \mgI\ near these velocities is 
not adequate to detect the scaled \mgII profile, so the velocity ranges
of absorption \mgI and \mgII are consistent with being the same.

\paragraph{FSC2349+24:}

The FSC2349+24 spectrum shows strong \mgII\ emission.
The 2796 emission fills in the  bluest portion of the 2803
line profile. The 2796 trough reaches -650\kms. We compare 
the \mgII trough to \mgI without any scaling. The SNR is not
adequte to detect the \mgI trough at outflow speeds greater
than -500\kms.


\begin{figure*}[h]
\centering
\includegraphics[scale=0.9,angle=-90,clip=true]{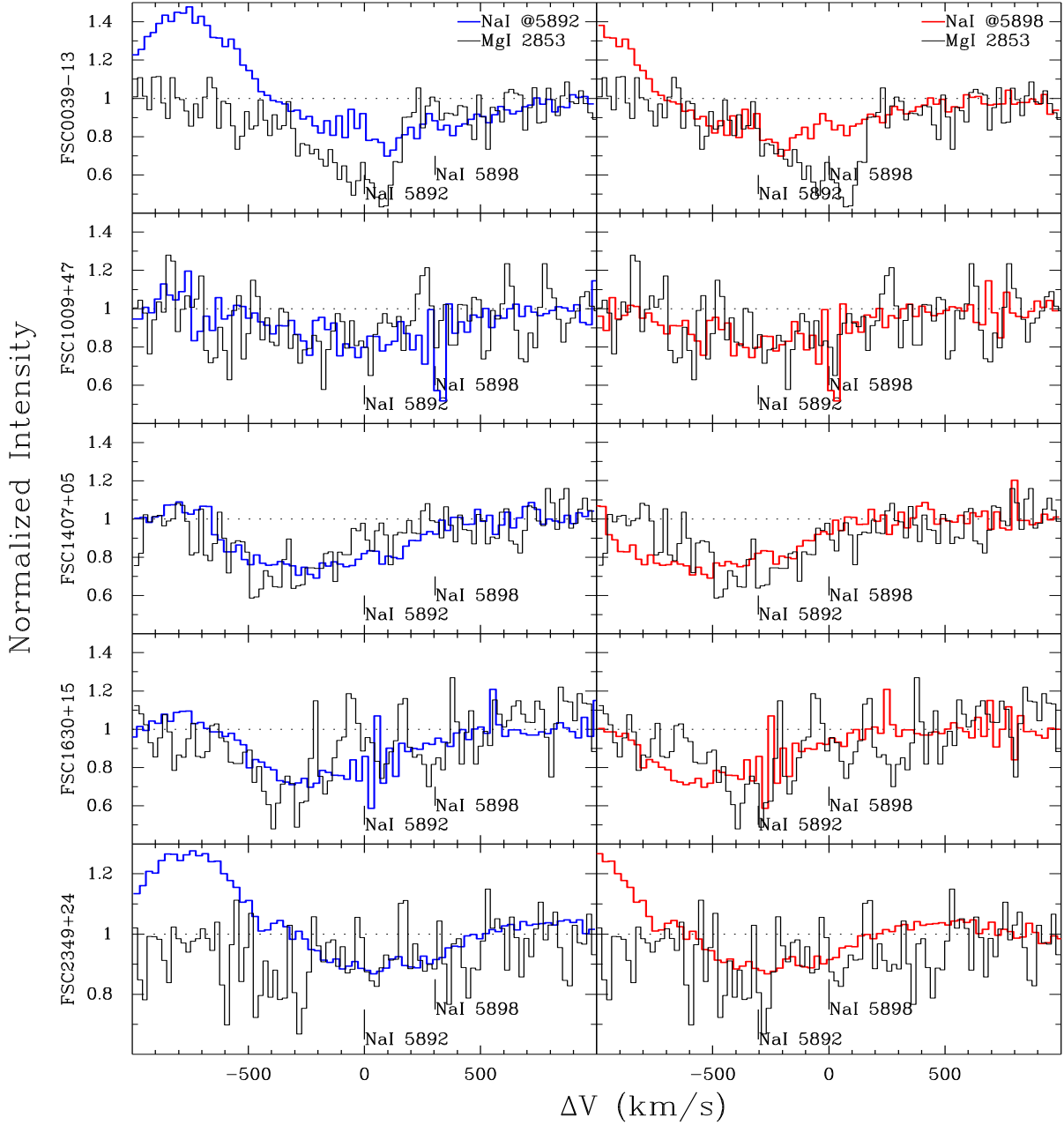}
\caption{Comparison of \nad and \mgI absorption line troughs in 5 ULIRGs.
{\it Left:} The blue side of the \nad trough, dominated by the \nad 5892 transition,
compared to the \mgI trough.
{\it Right:} The red side of the \nad trough, dominated by the \nad 5898 transition, 
compared to the \mgI trough.
}
\label{fig:nai_mgi} \end{figure*}

\subsubsection{The MgI Absorption Trough as a Template for the Blended \nad Trough} \label{sec:nai}

Since we detect these ULIRG outflows in neutral $Na$, the spectral energy distribution incident on 
the outflow likely presents a Lyman edge. (See the quantitative discussion in \S 2.3 of Murray \et 2007).
Stellar spectra have large intrinsic Lyman edges, so it follows that the starburst, rather than AGN,
likely determines the $Na$ ionization of the outflow. We expect a similar situation for $Mg$, which has
an ionization potential just 2.5~eV higher than that of $Na$ (7.6~eV vs. 5.1~eV). The patchy
distribution of dust in ULIRGs, however, leaves open the possibility that the morphological
appearance of the stellar component in the near-UV differs considerably from that at 6000\AA.
Absorption troughs present in these two bandpasses need not sample the same region of the outflow.

Before comparing the near-UV and optical trough shapes for the first time, it is useful to 
consider the relative optical depths in each transition should the absorption occur in
the same clouds. Magnesium is 19 times more abundant than Na but only four 
times more depleted than Na.\footnote{
          Depletion quoted for cool interstellar medium of the Galaxy (Savage \& Sembach 1996).} 
Singly ionized magnesium, \mgII, is the dominant ion of ${\rm Mg}$ over a large range of 
density and temperature. In contrast, the Murray \et (2007) ionization
models predict a larger variation in $\chi(\nad)$ from object to object, allowing a wide
range of possibilities for the relative optical depth of \mgI and \nad. 
For comparable Mg and Na ionization fractions, i.e. $\chi(\nad) \approx \chi(\mgII)$ where $\chi(X^{+n}) 
\equiv N(X^{+n})/ N(X)$, the \nad 5892 optical depth and that of the weaker \nad 5898 line will 
be lower than the \mgI 2853 optical depth.

The blending of the \nad 5892, 98 lines complicates the direct, model-independent
comparison. In \fig~\ref{fig:nai_mgi}, we use the \mgI profile as a template for both 
the \nad 5892 (left column) and 5898 (right column) absorption troughs. In three of the ULIRGs,
the depth of the \nad and \mgI absorption troughs are similar at minimum intensity. The 
two ULIRG spectra with strong HeI~5877 emission present shallower troughs in \nad relative
to \mgI, which we explain in part by emission filling of the \nad troughs. We take account of
the HeI 5877 emission filling in Section~\ref{sec:chi2} and estimate comparable maximum
outflow speeds in \nad,  \mgI, and \mgII.  Surprisingly, the modeling in Section~\ref{sec:chi2} 
demonstrates that the kinematics of the \nad absorbing gas are not distinghuishable
from that in the low-ionization UV transitions.

\paragraph{FSC0039-13:}
In the FSC~0039-13 spectrum, we detect \nad absorption up to -320\kms. Emission filling
from He~I plausibly explains why the \mgI and \mgII absorption troughs extend to higher 
velocity. The \mgI trough is nearly twice as deep as the bluer portion of the \nad 
trough (left panel) and three times deeper than the redder portion of the \nad trough 
(right panel). The HeI emission contributes to the shallowness of the blue half of the
\nad trough (relative to \mgI), but emission filling fails to expalin the low intensity 
of the red half of the \nad trough over 1000\kms from the HeI line. The deepest part of 
the \mgI profile at +90\kms matches the minimum in the \nad trough. At the wavelength 
corresponding to the same Doppler shift in the \nad\ 5892 transition, we find another 
local minimum in the blended, absorption trough. We postpone discussion of the relative intensities 
of the \nad 5892 and 5898 absorption troughs to Section~\ref{sec:chi2} because both transitions contribute to the 
absorption at minimum intensity.

\paragraph{FSC1009+47:}
In the FSC1009+47 spectrum, the \nad absorption trough extends from roughly the systemic 
velocity to about -450\kms.  At the coarse level of comparison allowed by the poor spectral
SNR, the shapes of \nad 5892 and \mgI profiles match without any scaling. Better SNR at
\mgII shows the outflow reaches -800\kms.

\paragraph{FSC1407+05:}
The \nad absorption is detected to -650\kms, close to (within one resolution element of)
 the maximum absorption velocity detected in \mgI and \mgII. The \nad and \mgI absorption 
trough shapes are very similar, with the latter being slightly deeper.

\paragraph{FSC1630+15:}
In FSC1630+15, the \mgI profile provides a good description
of the \nad\ absorption trough with no scaling
of the \nad 5892 or 5898 lines. Absorption is detected in both
troughs to -600\kms.
The SNR's of the \nad and \mgI spectra do not allow detection up to the
maximum velocity detected in \mgII.

\paragraph{FSC2349+24:}
In FSC2349+24, HeI emission
fills part the blue-end, $v < -250$\kms, of the \nad absorption trough.
The \mgI and \nad troughs have similar intensity at other velocities.

\subsubsection{Relative Strength of the \feII and \mgII Absorption Troughs}

The oscillator strength of \feII 2600 is about 3.5 times higher than that of \feII 2587. 
The similar shape of the two blueshifted \feII troughs requires even the \feII 2587 transition 
to be optically thick at the highest velocities measured in the outflow.
Iron has only a slightly lower cosmic abundance than Mg. Using cosmic abundance 
ratios, the optical depth of \feII 2600 will be 57\% that of \mgII 2803 for 
comparable depletion and ionization corrections. We directly compared the shape of the
\feII 2600, \feII 2587,  \mgII 2796,  and \mgII 2803  absorption troughs below,
omitting the two objects with low SNR near \feII. We find similar intensity
in all four troughs at comparable velocity.
Section~\ref{sec:chi2} presents quantitative, but model dependent, fits.

\paragraph{FSC0039-13:}

From -600\kms to +200\kms, $I_B \approx I_R$ for the \feII 2587, 2600 troughs.
The \feII 2587 trough does not present the dip seen in \feII 2600 from -800 to -600\kms.
We do not detect the  \mnII~2576 line, so absorption from the weaker \mnII~2594 transition
does not produce the blue dip in \feII 2600. We also exclude blending with excited, photospheric 
lines based on the location of such lines in the stellar spectra shown in 
Figure~\ref{fig:uvblue_feii}. We attribute the discrepancies in
the \feII profile shape over a small velocity range to a combination
of statistical and systematic errors in the data, where continuum fitting uncertainties
dominate the systematic error in this spectral region.
The \feII 2600 trough is identical to the \mgII 
trough (to within the error bars) over the velocity range from
-600\kms to +400\kms.  The \feII 2587 trough varies in lock step 
with the \mgII trough from -900\kms to +200\kms.  We conclude that
the \feII and \mgII absorption troughs sample gas with similar
kinematics.

\paragraph{FSC1009+47:}

The \mgII profile for FSC1009+47  provides a good description of 
the blue wing of the \feII 2600 trough with no scaling. The
\feII 2587 trough is slightly shallower.

\paragraph{FSC1407+05:}

The blueshifted wing of the FSC1407+05 \feII spectra at $v < -200$\kms 
shares the shape of the \mgII profile, but the \feII troughs are slightly
shallower (amplitude reduced by 1.3 in 2600 and 1.6 in 2587).
Both \feII troughs lie well below the \mgII
troughs from -100\kms to +100\kms. The \mgII 2803 profile presents
a weak, but significant, emission bump from 0 to +200\kms.  No \feII emission
is detected, so the \feII absorption troughs will look broader than 
the \mgII troughs. The \mgII absorption trough may share the kinematics 
of the low-velocity  \feII trough but be filled in by emission.

\subsection{Parametric Descriptions of the Absorption Troughs} \label{sec:parametric}

In the spectral regions where the \mgII 2796, 2803 absorption troughs 
can be directly compared, their intensities are nearly equal, $I_B \approx I_R$. 
Since the galaxy (continuum) likely subtends a large angle relative to that of an 
individual shell fragment or cloud, we might expect variation in the properties of
the absorbing material as a function of the spatial and/or velocity coordinates in
the outflow.  A number of methods have been developed to parameterize
the relative geometry of the outflow and continuum source. We apply the most common
method to the absorption troughs from FSC0039-13 and FSC1630+15, which show no sign of 
emission filling and little blending of the $\lambda 2796$ and $\lambda 2803$ lines. 
We then explain why two other parameterizations fail to describe the distribution of 
outflowing gas. 


\begin{figure}[h]
\centering
\includegraphics[scale=0.35,angle=-90,clip=true]{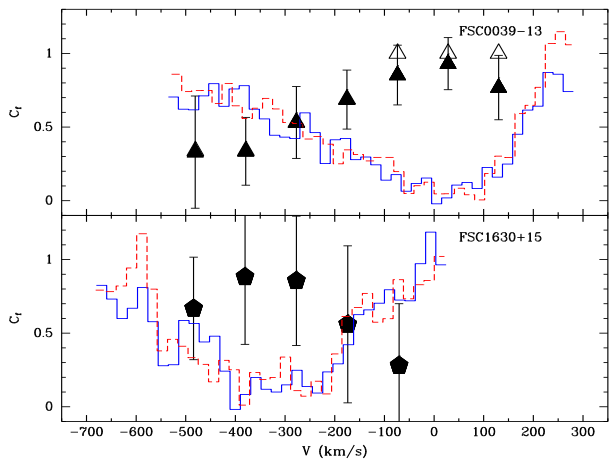}
\caption{Covering fraction vs. \mgII velocity. We binned to the spectral 
resolution and computed $C_f(v)$ from Equations~\ref{eqn:cv}.
For FSC0039-13, the open symbols show the effect of removing a
symmetric, zero-velocity component.
For reference, the solid (blue) line and dashed (red) line show the
the $\lambda 2796$ and $\lambda 2803$ \mgII absorption troughs.
We find that the covering fraction largely determines the shape of the absorption
trough. The troughs are deeper
at velocities where low-ionization gas covers a higher fraction of the continuum.
}
\label{fig:ctfit} \end{figure}

\subsubsection{Pure Partial-Covering Model} \label{sec:ppcmodel}

An approach used extensively for quasar winds (Arav \et 2001; Arav \et 2005; Arav \et 2008)
is to assume that a fraction $1 - C_f(v)$ of the area
of the continuum source is free of absorption at velocity $v$; and the
optical depth distribution in front of the covered part of the source is
constant. For this parameterization, the line intensities are
given by
\begin{eqnarray} 
I_R(v) / I_0 = 1 - C_f(v) + C_f e^{-\tau(v)} 
 \end{eqnarray}
\begin{eqnarray} \label{eqn:ib}
I_B(v) / I_0= 1 - C_f(v) + C_f e^{-2\tau(v)},
 \end{eqnarray}
where $\tau(v)$ is the optical depth at velocity $v$ in the weaker line, here
\mgII $\lambda 2803$. The coefficient in the argument of the exponential in Eqn.~\ref{eqn:ib} depends
on the relative oscillator strengths and wavelengths of the doublet transitions.
We binned the blue and red profiles to the spectral 
resolution; and then computed the covering fraction
\begin{eqnarray}
C(v) = \frac{I_R^2 - 2 I_R + 1}{I_B - 2 I_R + 1},
\label{eqn:cv} \end{eqnarray}
where the continuum has been normalized to unity, $I_0 = 1$.
At velocities where measurement errors produced unphysical line itensities,
$I_R < I_B$ or $I_R^2 > I_B$, we set the covering fraction to
 $C(v) = 1 - I_R(v)$ or $C(v) = 1.0$, respectively.
By substitution, the optical depth is
\begin{eqnarray}
\tau(v) = ln \left ( \frac{C(v)}{I_R(v) + C(v) - 1}  \right ). 
\label{eqn:tauv}  \end{eqnarray}
Where statistical fluctuations result in
$I_R < I_B$, measurement of the optical depth is 
not constrained. At velocities where  $I_R \approx I_B$, both transitions
must be optically thick; and this method cannot discriminate an optical depth of
3 from 30 in galaxy spectra of typical quality.  The limited spectral resolution and low
SNR of these data leave open the possibility of substantial systematic
errors in our best estimate $\tau(v)$. We will refer to this method as the pure 
partial covering scenario with  velocity-dependent covering fraction.

Figure~\ref{fig:ctfit} shows the covering fraction of low-ionization
absorption derived this way varies significantly across the trough.
Our analysis necessarily neglects components narrower than 100\kms,
which may contribute to the total gas column but are unconstrained by our spectra. 
The covering fraction is near unity only where the trough is black.
In the FSC0039-13 spectrum, this maximum coverage is near the systemic
velocity regardless of whether we remove a symmetric, zero-velocity
absorption component before computing $C(v)$. In FSC1630+15, the velocity
of the low-ionization gas with the highest covering fraction is between
300 and 400\kms. The covering fraction falls steadily with increasing
outflow speed in FSC0039-13, reaching  35\% at 500\kms. Similarly,
away from the velocity where the absorption trough presents minimum
intensity, $C_f(v)$ declines in FSC1630+15. 
The residual intensity correlates with the derived covering fraction.
We conclude that the shape of the absorption trough, $I(v)$, is strongly 
influened by the velocity dependence of the covering fraction of low-ionization
gas in the outflow.

\begin{figure}[h]
\centering
\includegraphics[scale=0.35,angle=-90,clip=true]{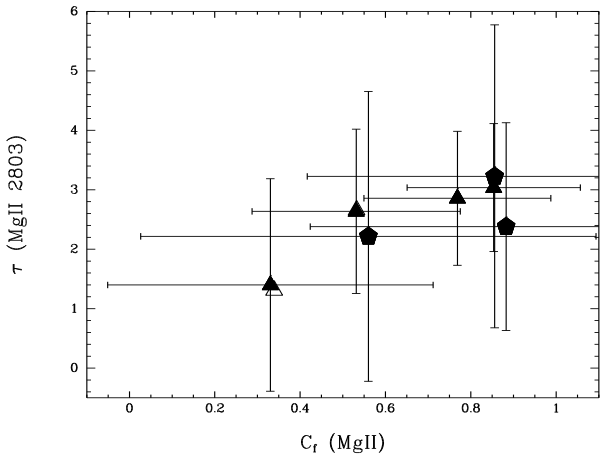}
\caption{Optical depth of the \mgII $\lambda 2803$
line vs. \mgII covering fraction. We binned the \mgII absorption troughs
to the spectral resolution and then computed $C_f(v)$ and $\tau(v)$ from 
Equations~\ref{eqn:cv} and ~\ref{eqn:tauv}. Triangles and pentagons represent
FSC0039-13 and  FSC1630+15, respectively.   Where statistical errors led to
unphysical intensity ratios, $I_B > I_R$, no upper bound on  the optical 
depth could be obtained; and no results are shown.
Removal of a symmetric, zero-velocity 
component in FSC0039-13 assigns zero covering fraction and optical depth to
the outflow  near the systemic velocity; but the fit to the outflow
portion of the trough barely changes (open symbols). The data allow, but
do not require, regions of the absorption trough with higher covering fraction 
to have higher optical depth.
}
\label{fig:tau} \end{figure}

Figure~\ref{fig:tau} compares the \mgII $\lambda 2803$ optical depth, $\tau(v)$, to $C_f(v)$.
The low-ionization gas at velocity $v \sim -600$\kms is apparently optically thick 
in the MgII transitions. At higher speeds, the blending of the blue wing of 
the \mgII 2803 trough with the \mgII 2796 line prevents a unique comparison of 
$I_R(v)$ to $I_B(v)$. The solution to Eqn.~\ref{eqn:tauv} for for FSC0039-13 and 
FSC1630+15 indicates that $\tau(v)$ increases by  a factor of $\sim 2 $ and 1.5, 
respectively, as the covering fraction grows by a similar factor.  Such a trend is
physically plausible, even probable, but is not required by the data. The
method only returns a lower limit on optical depth, $\tau \sgreat 3$. In our
spectra, optical depths of 3, 30, and 300 yield essentially identical
line profiles; in contrast, damping wings (not seen) would distinguish $\tau \sgreat 10^4$.
The lower limits on the optical depth, $\tau(v)$, determines the lower limit 
on the ionic column density in the ouflow at a given velocity. 
Applying Eqn.~1 from Arav \et (2001) to the optical depth in the
\mgII $\lambda 2803$ transition, 
\begin{eqnarray}
N [{\rm cm}^{-2}] = \frac{3.7679 \times 10^{14}}{\lambda_0[{\rm \AA}] f} \int \tau(v) dv,
\end{eqnarray}
we estimate total ionic column
densities of $N(\mgII) > 6.4 \times 10^{14}$\col\ 
and $ > 4.28 \times 10^{14}$\col\ toward FSC0039-13 and FSC1630+15,
respectively.


\subsubsection{Inhomogeneous Absorption}

The pure partial covering scenario with velocity-dependent covering fraction is just
one particular two-parameter ($C_f$, $\tau$) description of the relative intensities
of the doublet troughs. With just two absorption troughs to fit, other two-parameter 
models would also exactly fit the data. A real test of the model requires spectra covering
additional transitions from the same ion, which is not possible with our data. We can,
however, calculate parameter values for other models and examine how they change 
our view of the distribution of low-ionization gas in the outflow.

One such model is inhomogeneous absorption across the source.
This is appealing for AGN outflows because sharp edges in the spatial 
distribution of absorbing gas appear unrealistic because the  gas lies
at distances thousands of times greater than the size of the AGN emission source (de Kool \et 2002).
Arav \et (2005) present a simple model, where the optical depth distribution across a
uniform surface brightness source varies as $\tau(v,x) = \tau_0(v) x^a$, where $x$ describes
the location (in the plane of the sky) across the projected area of the source;  and $x$ is 
confined to the range [0,1]. To produce similar intensities in the \mgII $\lambda 2796$ and 
$\lambda 2803$ lines, they require the function $\tau(x)$ be steep, e.g. $a \sgreat 8$. 
A high value of the exponent works because it yields a factor of two change in the optical 
depth over a relatively small number of spatial elements, $x$, resulting in $I(2\tau) \approx I(\tau)$.
For even moderately large values of $\tau_0$, the optical depth will be low over at least a few 
tenths of the source area, and the absorption troughs will not be black.

In the ULIRG outflows, $I_R \approx I_B$ over a wide velocity range; but the \mgII troughs are 
nearly black over a smaller velocity range. The inhomogeneous absorber model requires unphysically 
large optical depths to reproduce these \mgII absorption troughs.
Starburst outflows also do not present the problem with pure-partial 
covering as do AGN outflows. The rotation observed across some low-ionization outflows suggest
the outflowing gas lies within a few kpc of the galaxy (Martin 2006).
Hence, we find no reason to adopt an inhomogeneous-absorption model for ULRIG outflows.


\begin{figure*}[h]
\centering
\includegraphics[scale=0.85,angle=-90,clip=true]{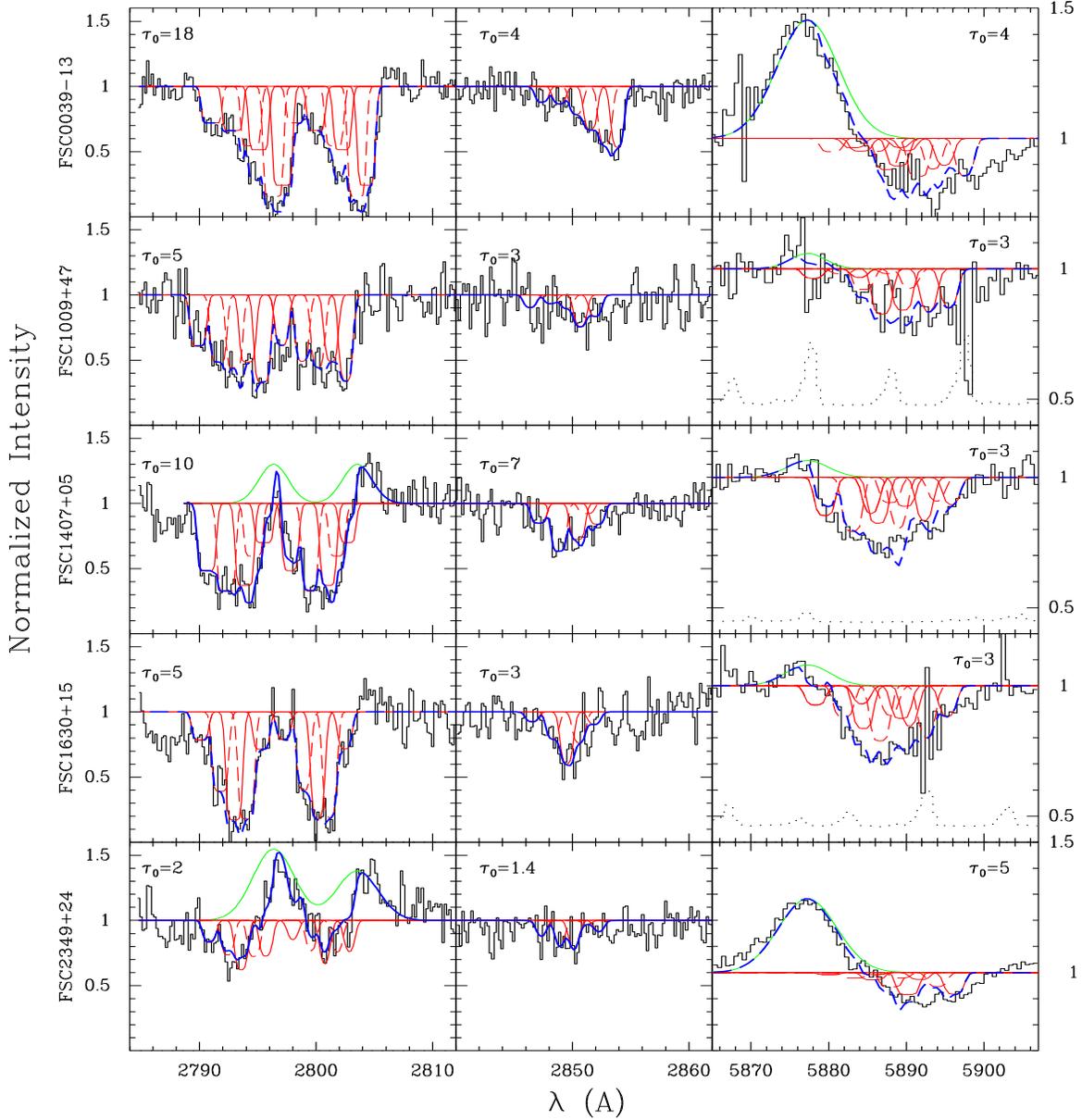} 
\caption{Fitted covering fraction of absorbing gas in each velocity component (red)
and, when required, emission lines (green).
{\it Left Column:} Fitted \mgII absorption troughs. All velocity components
have the Doppler parameter fixed at $b = 50$\kms and an optical depth at line
center of $\tau_0$. 
{\it Middle Column:} Velocity components fitted to the \mgI absorption troughs.
The Doppler shift and $b$ of each component matches the corresponding \mgII component,
and we fitted the velocity-dependent covering fraction $C_f(v)$ and minimum $\tau_0$.
{\it Right Column:} Velocity components fitted to blended \nad doublet trough. The
Doppler shift and $b$ value of each component matches the corresponding \mgII component;
we fit a velocity-dependent covering fraction $C_f(v)$ and the minimum $\tau_0$. 
Dotted line shows the relative intensity of sky emission; an
additive constant of 0.4 has been applied to the normalized sky intensity 
for display purposes. Although the fitted models are not unique; they demonstrate that
all three line profiles can be described with the same kinematic components. 
}
\label{fig:all_fits} \end{figure*}

\subsection{Fitting Velocity Components to the Absorption Troughs}  \label{sec:chi2}

The direct method of the previous section cannot be applied to the absorption troughs
in which the doublet transitions blend together and/or the absorption trough blends with
one or more emission lines. A quantitative description of these absorption troughs requires
fitting the parameters of functions describing each spectral component.

A covering fraction,  Doppler parameter, optical depth at line center, and Doppler shift
describe the attenuation of the continuum by one component, or cloud, at every
wavlength, $I_j(\lambda) / I_0 = 1 - C_j + C_j e^{-\tau_j(\lambda)}$. When the absorption
trough is a blended doublet, the optical depth at a given wavelength may include significant
contributions from each transition, and we add the doublet optical depths at that wavelength.
At any particular velocity, the two transitions of the doublet must have the same $C_f$, $b$, 
and Doppler shift; and atomic physics fixes the ratio of their optical depths.
At the systemic velocity of the HeI~5877 and \mgII emission lines, we fitted the 
amplitude of Gaussian intensity profile with velocity width matched to that measured for \Ha.

The relative intensities of the \mgII and \feII doublet troughs constrain
these transitions to be optically thick, setting a firm lower limit on the optical depth, 
$\tau_0 \sgreat 3$, at line center. The blue skew of the absorption troughs requires at 
least two velocity components. When we fitted such models, however, we found a number of 
parameter degeneracies.  We could describe 
the velocity width of each trough by adding additional velocity components, increasing the 
Doppler parameter, or further increasing the optical depth of individual components.  For
example, $\tau_0$ can be made arbitrarily large by simply allowing the Doppler parameter
$b$ to shrink; and the product $b \tau_0$ of a component can be made smaller (or larger) 
by adding (or removing) additional velocity components. Previous studies (Rupke \et 2005a;
Martin 2005, 2006) chose the minimum number of velocity components required to describe
the \nad absorption troughs.

Simulations of winds show that a single sightline intersects multiple shell fragments (Fujita \et 2009). 
And high resolution spectra of starburst galaxies resolve some broad absorption troughs 
into components (Schwartz \& Martin 2004). With this picture in mind, we assume a sightline intersects
a number of velocity components, where these structures partially overlap spatially. To allow for the 
possibility that some of the sightlines within the beam intersect different structures, we
adopt the partial overlap model introduced by Rupke \et (2005a). Each of the $n$ successive velocity 
components attenuates the continuum (and the preceeding components) such that
$I(v) = \prod_{j=1}^{n} I_j(v)$. A Maxwellian distribution provides a reasonable description of
the atomic velocities within each velocity component, so a Gaussian function models the optical depth
distribution, $\tau(\lambda) = \tau_0 e^{-(\lambda - \lambda_0)^2/(\lambda_0 b/c)^2}$, of each component. 
The covering fraction is constant across a given component but varies from one component to the next. The Doppler 
parameter, $b$, describes the turbulent motion along the line-of-sight in a given gas cloud.
For purposes of illustration, we fix the Doppler parameter at $b \equiv 50$\kms\ for all 
velocity components.   For a typical minimum value of $\tau_0 \sim 3$, this choice yields a line width 
at half  the trough depth around 100\kms, consistent with the spectral resolution.  We add velocity 
components until the fit statistic stops improving or a component becomes weaker than the detection limit. 
With this Doppler parameter, we need 6 velocity components 
to fit the \mgII absorption trough in FSC~0039-13 and 5 velocity components for the other 4 galaxies.

Allowing $\tau_0$ to vary among velocity components revealed no obvious trends with outflow 
velocity. The optical depth at the center of each component must be greater than three for the 
doublet transitions, but the probability distribution is very asymmetric with a significant tail 
reaching values $\sim 50$ times higher than the minimum value. (See the Monte Carlo approach described 
by Sato \et (2009) for describing uncertainties in optical depth.) We obtained an equally good
fit statistic when we required a single value of $\tau_0$ describe all the velocity components.
This assumption of a velocity-independent optical depth is our most dubious prior, and we discuss
how to improve upon it in Section~\ref{sec:mdot}. {\it We adopt this approach for the purposes of comparing the 
velocity-dependence of the gas covering fraction among objects and among transitions. For optically-thick
velocity components, the exact value of $\tau_0(v)$ has little effect on $C_f(v)$.
Our approach assumes the velocity-depdendence of the covering fraction determines the trough shape. }

\begin{figure*}[h]
\centering
\includegraphics[scale=0.7,angle=-90,clip=true]{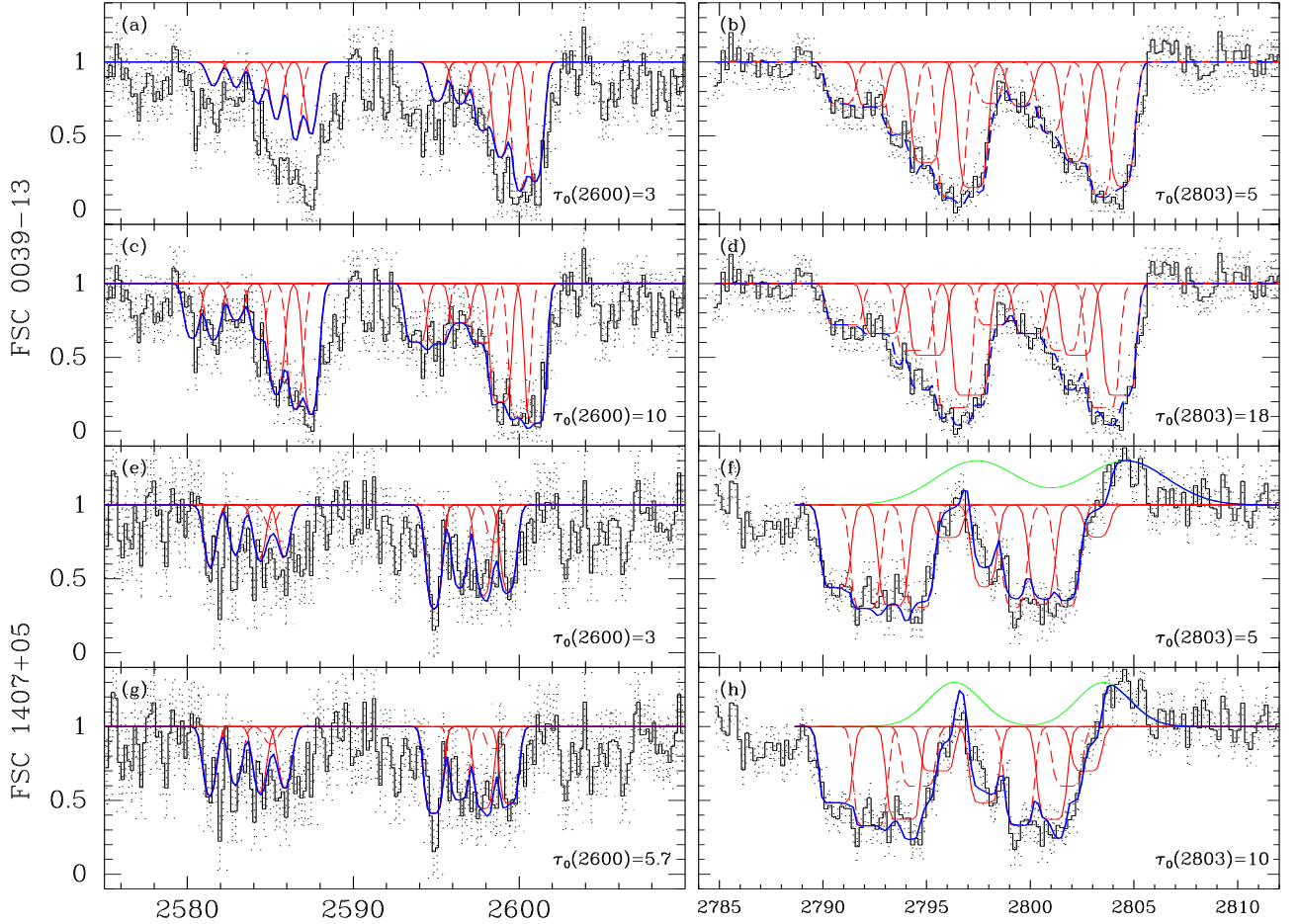}  
\caption{Constraints on optical depth from comparison of \feII and \mgII
absorption troughs. The \feII model in the left panel has $\tau_0(\feII 2600) = 
0.566 \tau_0(\mgII 2803)$.  The top and bottom two panels show FSC~0039-13 and
FSC~1407+05, respectively.
(a,b) {\it FSC~0039-13.}
Assumes the minimum \mgII optical depth that fits \mgII 2795 {\em and} 2803,
i.e. $\tau_0(\mgII 2803) = 5$. With $\tau_0(\feII 2600) = 3$, the \feII 2587 lines
are not nearly strong enough.
(c,d) {\it FSC~0039-13.} Fitting the \feII 2587 trough requires  $\tau_0(\feII 2600)$ of at least 10,
even when a 7th velocity component is included. Increasing  $\tau_0(\mgII 2803) $ to
17.6 is not inconsistent with the \mgII troughs.
(e,f) {\it FSC~1407+05.}
Assumes the minimum \mgII optical depth that fits \mgII 2795 {\em and} 2803,
i.e. $\tau_0(\mgII 2803) = 5$. With $\tau_0(\feII 2600) = 3$, the \feII 2587 lines
are not quite strong enough. The green line shows the fitted emission component.
(g,h) {\it FSC~1407+05.}
Fitting the \feII 2587 trough requires  $\tau_0(\feII 2600)$ of at least 5.7,
even when a 7th velocity component is included. Increasing  $\tau_0(\mgII 2803) $ to
10 is not inconsistent with the \mgII troughs.
{\it These results suggest the \mgII optical depth is likely a factor of $\sim 2$ to 3 larger
than the minimum required by the data.}
}
\label{fig:all_feii} \end{figure*}

We first fit $C_f(v)$ and the minimum value of the optical depth to the \mgII absorption troughs.
We found that the same kinematic components described the \feII and \mgI absorption troughs well. In
Section~\ref{sec:tau0}, we show that the minimum optical depth in the weak lines raises our best estimate
of the minimum optical depth in \mgII. We test the hypothesis these velocity components also describe
the blended, \nad absorption troughs and compare covering fractions among species. Table~\ref{tab:fit}
summarizes the fitted parameters. Figure~\ref{fig:all_fits} shows the resulting fits for \mgII, \mgI, and \nad. 
We plot \feII separately in \fig~\ref{fig:all_feii}.


\subsubsection{Fitted Optical Depth} \label{sec:tau0}

For \mgII, we can compare the ionic column inferred from the fitted partial-overlap model to those
estimated in Section~\ref{sec:ppcmodel}. With $b \equiv 50$\kms, we construct the same ionic columns by setting
$\tau_0(\mgII 2803) = 2.75$, which implies an integrated optical depth across each velocity component
of $\tau  = \sqrt{\pi} c^{-1} \tau_0 b \lambda_0 =$ 2.3\AA. The implied column densities,
\begin{eqnarray}
N[{\rm cm^{-2}}] = \frac{1.13 \times 10^{20} \tau_{tot}[{\rm \AA}]}{\lambda_o[{\rm \AA}]^2 f},
\end{eqnarray}
add up to $N(\mgII) = 6.4 \times 10^{14}$\col\ for the 6 components describing FSC~0039-13. 
This model has the minimum $\tau_0$ 
that comes close to describing the \mgII absorption trough given our prior on $b$. The reduced chi-squared
statistic improves significantly, from 1.425 to 1.143, when $\tau_0$ is increased to 5. The fit statistic 
does not get much better as $\tau_0$ is increased up to $\sim 50$, so we consider $\tau_0 \approx 5$ 
to be our lower limit when the \mgII absorption is considered independently. The pure 
partial-covering method also provided only a lower bound on $\tau_0(\mgII)$  due to saturation and low 
spectral SNR.

Based on inspection of the \feII absorption troughs, we argue that the actual \mgII 2803 optical depths 
are 2 to 3 times larger than the minimum required to fit the doublet.  For a cosmic 
abundance ratio of gas-phase iron to magnesium, the optical depth in the stronger iron line 
is  $\tau_0(\feII 2600) \sim 0.57 \tau_0(\mgII 2803)  \chi(\feII) / \chi(\mgII)$.\footnote
        {Although Fe and Mg have similar 1st and 2nd ionization potentials, the ionization
          equilibrium is complicated by the larger role of dielectronic recombinations for Fe
          (Shull \& Van Steenberg 1982). The Murray \et (2007) photoionization calculation for outflows 
          indicates that the largest difference in the \mgII and \feII ionization fractions occurs  
          at high gas density, $6 \times 10^5$\cm3, for the softest SEDs, when $\chi(\feII)$ can be 
          twice as large as $\chi(\mgII)$.}
For the weaker \feII line, $\tau_0(\feII 2587)$ will always be 3.5 times lower than $\tau_0(\feII 2600)$.
It follows that $\tau_0(\feII 2587)$ will be slightly less than unity if $\tau_0 (\mgII 2803) = 5$ {\it and}
if our estimate for the relative ionization correction, $\chi(\feII) \approx \chi(\mgII)$, is correct.
This model predicts the \feII(2587) absorption trough is much shallower than the \feII(2600) trough as
illustrated for FSC~0039-13 and FSC~1630+15 in panels~a and e, respectively,  of \fig~\ref{fig:all_feii}.

For FSC~0039-13,  this model spectrum is inconsistent with the shallow \feII(2587) trough, which requires
$\tau_0(\feII 2587) \sim 3$ or larger. As addressed further in Section~\ref{sec:density}, a difference
this large in ionization fraction is unlikely given predictions from photoionization modeling. We conclude
that the actual central optical depth in \mgII2803 is closer to 18 (than 5). Panel~d of \fig~\ref{fig:all_feii}
shows that the resulting \mgII fit is really indistinguishable from that presented in panel~b. We find the
fit to  \feII 2587 in panel~c preferable to that in panel~a.

For FSC~1630+15, the \feII 2587 absorption trough is slightly deeper than the prediction from
the minimum $\tau_0(\mgII 2803)$ model. Increasing $\tau_0(\mgII 2803)$ and $\tau_0(\feII 2587)$ to 10
and 1.6 provides an acceptable fit; larger optical depths make the \feII 2587 trough deeper than observed
in this case. We examined the UVBLUE spectra of synthesized stellar populations to determine if other
lines might be blended with the \feII 2600 line. We identified resonance transitions from
\mnII, marked in Fig.~\ref{fig:uvblue_feii}. 
The  \mnII 2594.499 transition does not cause the depression seen at slightly shorter wavelengths
because the stronger \mnII 2576.877 line is not detected. The SNR around the \feII 2587, 2600 lines is not
adequate for further comparison in the other objects.

Our observation of a single \mgI transition does not directly constrain the \mgI 2853 optical depth. 
The optical depth in \mgI is 
\begin{eqnarray}
\tau_0(\mgI 2853) = 6.1 \tau_0(\mgII 2803) \chi(\mgI) / \chi(\mgII). 
\end{eqnarray}
We argued that $\tau_0(\mgII)$ reaches at least $ \sim 10 - 18$, 
so the neutral Mg fraction must be less than 1 to 2 percent to produce an optically thin 
\mgI 2853 trough. It seems more likely to us that \mgI 2853 is optically thick. The 
covering fraction then determines the shape of the \mgI absorption trough, $I(v) \approx 1 - C_f(v)$.
Since covering fraction dictates both the \mgI and \mgII absorption trough shape, they have
a similar appearance, as shown quantitatively in the middle column of Figure~\ref{fig:all_fits}.
That the \mgII troughs are deeper indicates that only a fraction of
the cloud volume contains much neutral Mg.

The fitted models in \fig~\ref{fig:all_fits} assume equivalent kinematic components for \nad
and \mgII. For a cosmic abundance ratio, $N(Mg)/N(Na) = 19.07$, the \nad optical depth, 
$\tau_0(\nad 5898) = 0.115 \tau_0(\mgII 2803) \chi(\nad) /  \chi(\mgII)$, will be lower than
that of \mgII 2803 for similar ionization fraction. In Section~\ref{sec:density}, we argue that 
$\chi(\nad) /  \chi(\mgII)$ may be of order unity for soft spectral energy distributions (SED) 
with $L_{\nu}^{UV} / L_{\nu}^{FIR}  \sim 10^{-9}$ but will vary among outflows, dropping to
 $\chi(\nad) = 0.001$ for hard starburst SEDs.

When the weaker \nad line is optically thin, the fitted \nad models fall well short of the observed 
intensity in the redder half of the absorption trough. The models shown in \fig~\ref{fig:all_fits} have
$\tau_0(5898) = 3$ for 4 galaxies and 5 for FSC~2349+24. The ratio of these lower limits to those
for $\tau_0(\mgII 2803)$ in the same objects is 0.3 or more. Since the photoionization models, and
consideration of depletion factors, disfavor a situation with $\chi(\nad) /  \chi(\mgII) > 1$, the
moderate optical thickness required for \nad suggests $\tau_0(\mgII 2803)$ could easily be
a factor $\sim 2$ higher than that required by $\tau_0(\feII 2587)$. As discussed in Section~3.1.1,
however, the \nad and \mgII absorption troughs could probe physically distinct regions of the
outflow.


\begin{figure*}[h]
\centering
\includegraphics[scale=0.6,angle=-90,clip=true]{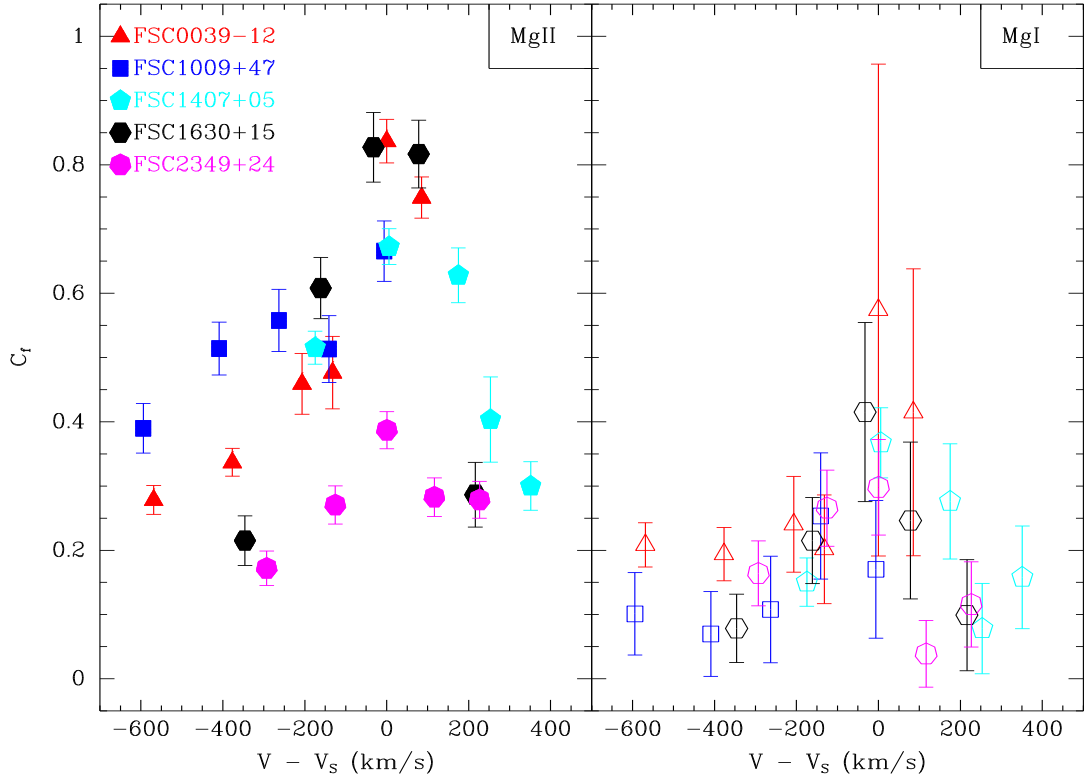}       
\caption{Fitted covering fractions to \mgII (solid symbols) and \mgI (open symbols) 
absorption troughs in five ULIRGs. Each component's velocity is shown relative to the 
velocity of the component having minimum \mgII intensity; this component may be
associated with the outflow velocity of a recently disrupted supershell. The shape of
the absorption trough is dictated by the velocity-depedence of the covering fraction 
because each velocity component is optically thick. This modeling approach quantifies
both the lower covering fraction of neutral Mg relative to singly-ionized Mg in the outflow
and the trend towards lower covering fractions at high velocity. 
}
\label{fig:cv_fit} \end{figure*}

\subsubsection{Constraints on Covering Fraction}

The attraction of forcing the same kinematic components is that we can directly compare
covering fraction of different species as a function of velocity, as summarized in Table~\ref{tab:fit}.  
The fitted $C_f$ values for the \feII velocity components come out similar to the values 
for \mgII. The maximum covering fraction in low-ionization gas occurs at minimum intensity by 
construction, and we again associate the velocity of maximum covering fraction with the speed of 
swept-up shells at breakout. The covering fraction decreases towards higher outflow velocity. 

Independent of any particular model, the shallow troughs of \mgI relative to \mgII require a 
lower covering factor for the former at every velocity. The $\chi_{\nu}^2$-fitting approach 
demonstrates that the shape of the \mgI absorption trough can be well described by the same
velocity components as the \mgII troughs and further quantifies the change in covering fraction. 
Figure~\ref{fig:cv_fit} compares the fitted covering fractions for \mgII and \mgI.  The covering 
fraction for neutral gas remains roughly  half that of the \mgII at all velocities. Given the
similar kinematics but constant offset in projected area, we suggest that the \mgI absorption 
originates in denser regions of larger structures traced by \mgII.

The fitted \nad covering fraction  is similar to that fitted to \mgI for each velocity component. 
Sodium and magnesium have similar first ionization potentials, 5.1~eV for Na, and 7.6~eV for Mg.  
In the FSC0039-13 spectrum, emission filling could explain shallow \nad 5892 trough, 
relative to \mgI; but the emission is not broad enough to account for the shallow \nad 5898 trough.
We suggested that the FSC~0039-13  sightline intersects more of the galactic, gas disk than 
our other observations; and the high inclination of this disk means we probe the swept-up shell 
where it has stalled in the disk.  The extra attenuation at high inclination would likely
produce larger than average differences between near-UV and optical continuum morhpology.
High-resolution imaging in these bands may yield further insight into why the covering
fraction of neutral, alkalai metals differs by almost a factor of two between the near-UV
and optical sightlines.  In more typical ULIRGs, the similar kinematics of \nad absorption 
and that of \mgI, suggest the \mgI velocity components and $C_f(v)$ might be applied as template 
for deblending the \nad absorption trough into its \nad 5892, 98 components.

\subsubsection{Minimum Mass-Loss Rate}  \label{sec:mdot}

In a spherical outflow geometry, the mass flux in the low-ionization 
outflow, $\dot{M}_c(R) = \Omega C_f(R) v(R) \rho_c(R) R^2$.
The mass column, $\bar{m} N_H(v) = \int_{R1}^{R2} \rho(r) dr$, measured for any velocity
carries a mass flux of $\dot{M}(v_i) = \Omega \bar{m} N_H(v_i) C_f(v_i) v_i R_1 R_2 / 
(R_2 - R_1)$, where the distance and radial depth of the component are model
dependent.
We expect most of the mass-loading in the starburst region and therefore that
$\dot{M}(r)$ be approximately constant throughout the flow. For this to happen,
the product $\tau_{0,i} C_{f,i} v_i$ needs to decrease as the radial distance to component $i$,
$r_i^{-1}$, increases.  Our
fitted values of covering fraction decline with increasing velocity, but
the product $v_i C_f(v_i)$  grows with increasing outflow velocity. Obtaining a constant
mass flux requires the central optical depth to decline with increasing
velocity.  If we choose the minimum optical depth allowed for the highest
velocity component, then one can construct the values $\tau_0(v)$ fro $v < max~v$
needed to raise the mass flux at lower velocities to the same outflow rate,  
i.e. $\dot{M}(v)$ equal to a constant. We obtain the equivalent answer
by estimating the mass flux from the highest velocity component. 

For FSC~0039-13, the velocity component at 571\kms carries a mass flux
of $\dot{M} \ge 6 \msunyr \chi(\mgII)^{-1} R / \Delta R (R / 10 {\rm ~kpc}) (\Omega / 4\pi)$,
where the lower limit comes directly from the requirement that $\tau_0(2587) \ge 2.7$.
The velocity component at 615\kms in the \feII absorption troughs of FSC~1407+05
requires $\dot{M} \approx 5 \msunyr \chi(\feII)^{-1} R / \Delta R (R / 10{\rm~kpc}) (\Omega / 4\pi)$.
In Section~\ref{sec:density}, we will argue that the ionization corrections for \mgII and
\feII are of order unity. The unknown distance to the absorbing gas and radial thickness
of the component leave order-of-magnitude uncertainty in the lower limit.

Comparison of cosmic abundances and atomic data indicates $\tau_0 = 3$
will yield the highest mass-loss rate for \nad 5898, with \feII 2587 
about 66\% as large. To make the implied mass columns agree, the minimum optical
depth in \feII 2587 would need to be $4.5 C_f(\nad) / C_f(\feII) \chi(\feII) / \chi(\nad)$,
or roughly $2.25 \chi(\feII) / \chi(\nad)$. In Section~\ref{sec:density}, we find the upper
bound on the correction to the minimum optical depth,  $1 / \chi(\nad)$ could reach $10^{3}$ for
those ULIRGs with relatively hard SEDs. The lower limits on the mass fluxes likely yield an 
accurate picture for some systems but underestimate the true mass-flux in low-ionization gas
by orders of magnitude in others.

\subsection{Gas Volume Density}   \label{sec:density}

The volume density of the low-ionization gas varies widely among outflow models.
Figures~4 and 5 of Murray \et (2007) illustrate the sensitivity of the \mgII and \nad
ionization fractions to the gas density. Since the saturated lines provide only lower 
limits on the column densities of \mgI and \mgII, we cannot directly estimate the ionization 
fraction, thereby  constraining the volume density. The null-detection of absorption from
collisionally-excited levels above the ground state provides a useful upper limit on the electron 
density, however. The most relevant excited transitions of \feIIe are marked in \fig~\ref{fig:uvblue_feii};
see Korista \et (2008) for a summary. We compute this upper limit and then discuss how it 
constrains ionization fraction in the outflow.

Above the critical density, the level populations approach their Boltzmann ratio, and we
can easily calculate the relative strengths of absorption lines from different energy levels.
The lowest energy level above the ground state, $E = 385$~cm$^{-1}$, has the lowest critical 
density. Our spectra cover the second strongest line from the multiplet with this lower 
energy level, \feIIe 2612.6542.

In the FSC0039-13 spectrum, we expect \feIIe 2612.6542 to be the strongest
excited line. We place an upper limit on the observed equivalent width using the
continuum SNR and assuming a line-width comparable to a spectral resolution element.
The $5\sigma$ upper limit on the corresponding rest-frame equivalent width is 0.40\AA.
For comparison, we also fitted a single-absorption-line model with $b \equiv 50$\kms and
$C_f \equiv 1$ to the spectrum. At the systemic velocity, we find $\tau_0 (\feIIe 2612) < 0.30$,
a value typical of this spectral bandpass. We obtain a more conservative limit if we
repeat this procedure where the largest continuum deviation occurs at -850\kms; and we
find $\tau_0 (\feIIe 2612) < 0.62$. This more conservative limit corresponds to a column 
density, $N(\feIIe) < 6.26 \times 10^{13}$\col\ per component.

The optical depth in the individual velocity components of the  weakest, detected 
resonance line is $\tau_0(\feII 2587) \ge 3$. Each of the 6 velocity components
had $b \equiv 50$\kms, so the lower limit on the column density in the ground state
is $N(\feII) > 5.61 \times 10^{14}$\col. Our measurements place a solid upper limit
on the relative column densities in the first-excited and ground states,
$N(\feIIe) / N(\feII)  < 0.11$, and likely less than 0.05. For densities well above the critical
density this ratio would be $\sim 0.75$ with a slight dependence on temperature. Using
the level population calculations from Figure~3 of Korista \et (2008), we
conservatively limit the electron density to $\log n_e < 3.5$ (or 3.4) for a
temperature of $1 - 1.5 \times 10^4$~K (or $5 \times 10^3$~K), respectively. The stronger
limit of $N(\feIIe) / N(\feII)  < 0.05$, which applies if the highest density gas is 
at low velocity, lowers $\log n_e$ to 3.1 (or 3.0), respectively, at $T = 1-1.5 \times 10^4$~K
(or $T = 500$~K).

The gas density is important for
understanding the relationship between the hot wind and the outflow observed
in UV-optical absorption lines. Photoionization equilibrium likely sets the 
temperature of the low-ionization gas at about $10^4$~K. The density of the cool, low-ionization
gas must be  $> 50$\cm3  to be in pressure equilibrium with the hot wind, where $P_h \sgreat 
(10^7 {\rm ~K})(0.05 {\rm ~cm}^{-3}) \sim 5 \times 10^5$~K\cm3.  
Outflows accelerated by radiation pressure on dust grains
do not require a hot wind at all (Murray \et 2005), allowing the low-ionization gas to have much
lower density. Our upper limit on the gas density does not challenge the multi-phase models.

Our result does eliminate the Murray \et (2007) ionization models with $n \sim 6 \times 10^4$\cm3 or 
larger (see their Figures 4 and 5).
We can confidently claim that \mgII is the dominant ionization state of Mg. For any reasonable
starburst spectral energy distribution (SED), $\chi(\mgII) \ge 0.7$; and the neutral fraction
is less than 30\%. The correction from the ionic columns of $N(\mgII)$ (and $N(\feII)$) to the
elemental columns will be relatively minor. This knowledge of the Mg ionization balance suggests the 
\mgI 2853 optical depth significantly exceeds its minimum value of 3. Obtaining the same total column 
of Mg from the  \mgI 2853 and \mgII 2803 components requires 
\begin{eqnarray}
\tau_0(2853) = 6.1 \frac{\chi(\mgI)}{\chi(\mgII)} \frac{C_f(\mgII)}{C_f(\mgI)} \tau_0(2803),
\end{eqnarray}
or $\tau_0(2853)$ up to $5.2 \tau_0(2803)$ when $\chi(\mgI) = 0.3$ and $C_f(\mgI) = 0.5 C_f(\mgII)$.

Eliminating very high density rules out a high neutral fraction of Na. Interpolating between
the curves in Figure~4 of Murray \et (2007) to a density of 1000\cm3, the \nad ionization 
fraction plumments from unity, when $L_{UV}/L_{IR} \le 10^{-9}$,  to $\chi(\nad) \sim 10^{-3}$ 
when the SED hardens to  $L_{UV}/L_{IR} \sim 10^{-4}$. In contrast, only at very high gas density does
the neutral Mg fraction change rapidly with spectral hardness. The ionization potential of Mg is 
2.5~eV higher than that of Na..

\subsection{Emission}

Our spectrum of FSC2349+24 very clearly shows \mgII emission. 
The emission line appears to be a bit redshifted, but we argue that
the \mgII absorption attenuates the blue side of the emission profile.
The only other spectrum that definitely presents \mgII is FSC1407+05.
The \mgII 2803 profile shows the emission from 0 to +250\kms. The 
corresponding \mgII 2796 line is less obvious because of \mgII 2803 
absorption at the same wavelength. We suggested that emission filling of the 
\mgII trough explains the deeper absorption in the \feII trough relative to 
the \mgII trough near systemic velocity.

Weiner \et (2009) discovered \mgII emission in a subset of $z \sim 1.4$ galaxies
with outflows and argued that the composite spectrum of the non-emission galaxies
presented weak emission. They demonstated the presence of the emission by fitting
the red-shifted portion of the \mgII 2803 absorption trough with a symmetric absorption 
component at the systemic velocity.  Removal of this component yielded a redshifted
emission component in \mgII 2796. This technique cannot be applied to four of five
galaxies in our sample because their \mgII absorption troughs present no 
absorption at $v > 0$\kms. Removing a symmetric, zero-velocity component fitted
to the redshifted absorption trough in FSC0039-13 reveals no significant 
emission excess in \mgII 2796.  

The origin of the \mgII emission is not completely clear, but we adopt
the hypothesis that the \mgII lines are excited by recombination. 
The two objects presenting emission are classified as Sey~2 on the basis of
their optical emission-line ratios, suggesting the ionizing spectrum is harder.
The absence of extended \mgII emission in the 2D spectra of FSC1407+05 and FSC2349+24
rules out a scattering origin from a galaxy-scale nebula.

The fitted profiles are shown in Figure~\ref{fig:all_fits}
In FSC1009+47, the red side of the \feII troughs are marginally lower than
those of \mgII, hinting at a hidden emission component in the latter. 
However, inclusion of an emission component in the model for any of the
other three galaxies does not improve the fit to the \mgII absorption trough. 
In fact, any emission component at the systemic velocity significantly degrades
the fit if the maximum intensity is more than $10\%$ of the continuum level.
We conclude that only two of the five ULIRGS present significant \mgII emission.
In contrast, an He~I~5876 emission improves the model for the \nad absorption trough in
all five galaxies. The line is strongest in FSC0039-13, which is the only object
classified as an HII galaxy in our sample.

\subsection{Dynamical Ages of Targets}

In the popular ``cool ULIRGs $\longrightarrow\ $ warm ULIRGs $\longrightarrow\ $ quasars''
evolutionary scenario, as suggested by Sanders \et (1988), one expects the AGN to
provide the increase in bolometric luminosity. Advanced mergers in the 1 Jy sample tend to 
have higher luminosity (Veilleux \et 2002). While our sample does contain rather luminous
ULIRGs, their 25-60\um colors, see Table~\ref{tab:sample}, are all red enough to be classified 
as cool ULIRGS, i.e. $f_{25}/f_{60} < 0.2$.

Following the  morphological classification scheme outlined in
Veilleux \et (2002), the most {\it advanced mergers} we observed are
FSC0039-13 and FSC1407+05. These are single nuclei systems with a 
compact morphology and little tidal structure. The dynamically younger
merger, FSC1009+47, presents tidal tails but the nuclei have coalesced. 
The tails extend to 29~kpc in FSC1009+47 (Veilleux \et 2002).  Our second 
spectrum for FSC1009+47 is extracted 44~kpc south (along the slit) of the
nucleus. Veilleux \et (2002) classify this feature as a prominent 
knot in a tidal arm rather than a second nucleus because it is
not detected in their K' image.
Pre-merger sources with separated nuclei are rare in the 1~Jy sample,
but we observed at least two such objects. FSC1630+15 is a close binary,
separation 4.4~kpc (Veilleux \et 2006); and  FSC2349+24 is a wide binary,
separation $14.5$ kpc. 

Among this small sample we found no evolutionary trends in low-ionization 
outflow properties.  
We compared the outflow properties along the sightlines to each nucleus
in the double systems. The outflow was always seen in each spectrum.

\section{DISCUSSION} \label{sec:discussion}

We detected outflows in 5 ULIRGs in ground-state absorption
from \nad, \mgI, \mgII, and \feII and illustrated the remarkably 
similar shapes of the absorption troughs in all these transitions. We 
then demonstrated that the velocity-dependence of the gas covering fraction
determines the trough shape.  Similarities in trough shape suggest these 
species reside in the same, low-ionization gas structures. Because
\mgII (and \feII) absorption covers a higher fraction of the continuum 
source, at a given outflow velocity,  than do neutral Mg and Na, we argue
that the absorbing clouds or filaments are not homogeneous.

The Doppler shift of the absorbing gas does not identify its
position along the line-of-sight. In particular, no consensus
had been reached previously as to whether the highest velocity material
detected resides close to the starburst region or at much larger radii.
A close connection likely exists between this neutral-atomic gas, outflowing dust, 
and large-scale molecular outflows. Images revealing the presence of dust 
and molecular gas in M82 out to a few ~kpc (Walter \et 2002; Hoopes \et 
2005; Veilleux \et 2009b) provide strong evidence for the survival (or 
continuous creation) of dense clouds despite ablation by the hot wind 
and evaporation, as described (for example) by Marcolini \et  (2005).

Along the outflow axis of M82, it remains unclear whether the denser material 
shares the same kinematics as the \Ha emission. The \Ha Doppler shift does
increase with increasing distance along the minor axis (Heckman \et 1990; 
Shopbell \& Bland-Hawthorne 1998; Martin 1998), consistent with acceleration.
However, we lack such well-resolved, position-velocity information for most 
nearby (and all high-redshift) galactic outflows. In this section, we describe
what the velocity-dependence of the gas covering fraction may imply about
the location of the low-ionization absorbing gas along the line-of-sight.
We begin by considering some pedantic dynamical models, starting
with the blowout of a superbubble as described by DeYoung \& Heckman (1994).

\subsection{Development of a Galactic Wind}

The thermalized energy from supernova explosions drives a shock front
through a galaxy, sweeping interstellar gas into a thin, radiating supershell
(Tenorio-Tagle \& Bodenheimer 1988; Shull 1993). While the growing shell remains
smaller than the pressure scale height of the interstellar medium, it plows
through an ambient medium of essentially uniform density; and the shell decelerates 
with velocity falling as $v \propto t^{-2/5}$, 
or equivalently $v \propto R^{-2/3}$,
for a  continuous injection of mechanical energy (Weaver \et 1977).
The shell mass grows linearly with the bubble volume at
this stage, and the mass column through the shell grows linearly with radius
\begin{eqnarray}
\bar{m} N(r) = 1/3 r \rho_0,
\end{eqnarray}
where $\rho_0$ is the average density of the ISM. During the supershell phase, 
the covering factor of the low-ionization gas will be unity 
provided the shell traps the ionization front.

Emission-line images of nearby, starburst galaxies show that supershells outgrow
their host galaxies. We expect the shell to accelerate when the density gradient becomes 
steeper than $\rho(r) \propto r^{-2}$ (McKee \& Ostriker 1988), a  
highly unstable situation in which a dense shell pushes on more rarefied gas.
Numerical simulations show that the shell breaks up due to hydrodynamic
instabilities at a few pressure scale heights (MacLow, McCray, \& 
Norman 1989). The evolution of the shell fragments has received relatively 
little attention, but they are clearly one source of low-ionization gas that 
will absorb continuum emission. Additional sources of low-ionization gas
include hydrodynamic instabilities induced by shear at the disk -- wind 
interface (Heckman \et 2000) and pre-existing, interstellar clouds over run 
by the supershell (Cooper \et 2008). 

A primary origin for the low-ionization outflows in shell fragments 
is appealing because it naturally explains the velocity offset of the \mgII 
absorption troughs from the systemic velocity. Due to the high central 
concentration of gas in ULIRGs, the pressure scale-height is several times 
smaller than the value $h_z \sim 100$~pc typical of normal galaxies; and
theblowout radius  $R_0 \sim 3 h_z \sim 200$~pc. By the time of
blowout, the mean speed of a shell has dropped to
\begin{eqnarray}
v (R) &=& 164 {\rm ~km~s}^{-1} 
\left( \frac{L_w}{ 7.08 \times 10^{43} {\rm ~erg~s}^{-1}} \right) ^{1/3} \nonumber \\
&& \left( \frac{10^3 {\rm ~cm}^{-3}}{ n_H}\right) ^{1/3} 
\left( \frac{200 {\rm ~pc}}{R}\right)^{2/3},
\end{eqnarray}
where the mass per H atom is $\bar{m} = 1.4 m_H$ in the ambient medium and $L_w$, the rate 
of  mechanical energy injection, has been scaled to a SFR representative of
ULIRGs, i.e. 100\msunyr.\footnote{ 
        Value of $L_w$ from SB99 continuous
        star formation model (Leitherer \et 1999)
        for 1\msunyr in 1 to 100\msun\ stars.}
A shell velocity $\sim 200$\kms can describe the Doppler shift of 
the deepest part of the \mgII absorption troughs in \fig~1b-1e.
A sightline at high inclination (i.e. close to the plane of
the gas disk) would intersect a lower velocity shell due to the higher average
gas density in the plane. We appeal to this scenario as a plausible explanation 
for the  kinematics of the \mgII absorption troughs in FSC0039-13, which are
deepest near the systemic velocity.

Numerical simulations of ULIRGs (e.g. Fujita \et 2009) confirm that shell velocities decline to 
200-300\kms by the time of blowout and suggest that, following blowout, the hot interior of the bubble 
accelerates outward creating a hot wind with terminal velocity  
$v_h \approx \sqrt{3} c_s \sim 940 T_7^{1/2}$\kms.  Fujita \et did not follow the 
shell fragments very far into the halo;  they simply assumed the {\it clouds}
coast outwards on ballistic trajectories. Whether or not
cosmic rays (Breitschwerdt 2008; Everett \et 2008; Socrates \et 2008), 
radiation pressure, and/or the ram pressure of the hot wind further accelerates
the low-ionization gas remains an important question.

Murray \et (2005) analytically modeled low-ionization outflows accelerated by the 
ram pressure of a hot wind and outflows accelerated by radiation pressure. It 
remains challenging to observationally distinguish not only these two types of 
momentum-driven outflows but also the energy-conserving, ballistic trajectories.
The former accelerate quickly and then coast. The latter coast for a large
distance before gradually decelerating. Factors favoring radiative-driving 
include an empirical correlation between outflow speed and escape velocity (Martin 2005), the
high dust content of ULIRGs, and the high luminosity of ULIRGs. Reasonable
objections include the small number of dwarf galaxies used in the outflow
speed correlation, the difficulty of coupling radiative momentum to the gas
in dust-poor, lower luminosity dwarf galaxies, and evidence for the presence
of hot winds in ULIRGs (Sciortino, V. \& Martin, C. L. in preparation).

For purposes of illustration, we consider acceleration by the ram pressure of
a hot wind following blowout. The shell radius, $R_0$, and velocity, $v_0$, at blowout 
set the initial conditions at the start of the  wind phase. In a spherical
outflow geometry, the ram pressure drops as the inverse-square of the radial distance,
so clouds accelerate over a relatively short spatial scale. The terminal velocity of any
shell fragment depends on its column density, with the low columns reaching, at most,
the hot wind velocity, $v_c \approx v_h$ (Martin 2005). {\it The absorption troughs presented 
in this paper require the covering fraction of shell fragments to decrease with increasing 
velocity. This situation can be achieved by tuning the distribution of cloud column densities.
The lowest column density fragments, which reach the highest terminal velocities, would need
to be relatively rare, covering less area than higher column density clouds. Alternatively,
the geometrical dilution of clouds offers a simpler way to achieve the desired result.} 
We demonstrate this largely geometrical effect in Section~\ref{sec:dilution}.
Further numerical work, beyond the scope of this paper, will be required to address the broader problem.

\subsection{Geometrical Interpretation of the Velocity-Dependent Covering Fraction} \label{sec:dilution}

To illustrate the importance of geometrical dilution, let the locus of fractures in a
shell define individual clouds. Suppose, for simplicity, these clouds all
have the same column density and intitial velocity, equal to that of the shell just before
blowout. If the area of a cloud does not change as it is moves outwards (think of a
brick-like cloud), then the covering fraction of clouds decreases as $C_f(R) = C_f(R_0) (R_0 / R)^2$.  

Most shell fragments are unlikely self-gravitating structures, however; and we
expect them to expand as they fly outwards due to the drop in ambient pressure. The sound
crossing time in a cloud is likely short enough for the cloud to maintain pressure
equilibrium with the hot wind, $P_c \approx P_h$. The clouds expand adiabatically, so
we can calculate their size at any outflow radius, $R$, and compare the cloud area
to the that of a solid shell, thereby estimating the covering fraction, $C_f(R)$.

Mass conservation in a steady-state, constant velocity hot wind requires the density
to fall as $\rho_h \propto r^{-2}$.  For an isothermal hot wind, the resulting increase
in the volume of low-ionization clouds is
\begin{eqnarray}
V_c(R) = \left( \frac{R}{R_0} \right)^{2/\gamma_c} V_c(R_0),
\end{eqnarray}
where $\gamma_c = 5/3$ for a monatomic, ideal gas. For roughly spherical clouds,
the increase in cloud volume is accompanied by an increase in cloud area,
$ A_c \propto V_c^{2/3} $.  Defining the covering fraction as
\begin{eqnarray}
\frac{C_f(R)}{C_f(R_0)} = \frac{A_C(R)}{4\pi R^2} \frac{4\pi R_0^2}{A_C(R_0)}, 
\end{eqnarray}
these relations yield a covering fraction,
\begin{eqnarray}
\frac{C_f(R)}{C_f(R_0)} = \left( \frac{R}{R_0} \right)^{4/3\gamma_c -2},
\end{eqnarray}
that falls as $R^{-1.2}$. 

If the hot wind cools adiabatically, then $T_h \propto 
R^{-4/3}$; and the pressure falls off faster with radius. The lower pressure
allows the clouds to expand faster than in the isothermal case. Repeating
the steps in the previous paragraph, we find
\begin{eqnarray}
\frac{C_f(R)}{C_f(R_0)} = \left( \frac{R}{R_0} \right)^{4  \gamma_h / 3\gamma_c -2}.
\end{eqnarray}
Assuming $\gamma = 5/3$ for both the hot wind and the clouds, we find
\begin{eqnarray}
\frac{C_f(R)}{C_f(R_0)} = \left( \frac{R}{R_0}\right)^{-2/3}.
\end{eqnarray}
The low-ionization clouds cannot expand quickly enough to
keep up with the geometrical dilution inherent to  spherical, outflow geometry. 
Their covering fraction must decrease with increasing outflow radius.

\subsection{Is Acceleration of the Low-Ionization Gas Required?}

The absorption trough measurements require lower $C_f$ at higher velocity. Geometrical
dilution produces lower covering fractions at larger outflow radii. Combining these
two results implies that the gas producing the highest velocity absorption resides at the 
largest distance from the starburst. While we find this argument illuminating, some
of the underlying assumptions should be examined.

First, the outflow geometry is uncertain.
It should be roughly spherical on spatial scales larger than the galaxy, but the 
outflow geometry may be better described as cylindrical at blowout. We know
the absorbing gas lies close to the gaseous disk 
in some ULIRG outflows because the outflows present rotation in \nad (Martin 2006).
In a version of the above model with cylindrical geometry, both the covering fraction and 
column density of 'brick-like' clouds become independent of height above the disk. 
The decrease in covering fraction would require other physical effects such as 
cloud ionization, evaporation, or ablation to become important at the higher outflow
velocities.

Second, blowout may yield a distribution of cloud column densities and velocities,
something three-dimensional numerical simulations may soon be able to address.  The 
fastest fragments -- whether determined by blowout, acceleration by the hot wind, or 
acceleration by the starburst radiation -- will reach the largest distances. Our
models indicate the low-ionization cloud reach their maximum distance long after the
starburst activity has ceased. It follows that if we observe such outflows during
the starburst phase, the highest velocity absorption will come from material at
the largest radii. Our underlying explanation for the shape of the absorption troughs
would remain geometrical dilution; however, acceleration of the low-ionization outflow
would not be required beyond blowout.

\subsection{Measurement of Terminal Velocities and Estimates of Spatial Extent}

Comparison of the \mgII, \mgI, and \nad absorption troughs indicates
the highest velocity gas sometimes escapes detection in \nad and \mgI.
For example, in FSC0039-13, we detect MgI and NaI absorption to 300\kms; but 
the MgII absorption demonstrates that the outflow persists to
500\kms. In FSC1407+05, we measure terminal velocities
of 350, 600, and 750\kms from \nad, \mgI, and \mgII, respectively.
We find our spectral signal-to-noise ratio insufficient to detect the higher 
velocity components of these outflows in \nad and \mgI. 

The \mgII lines 
allow more robust terminal velocity measurements for a number of reasons.  
First, the \mgII 2796, 2803 transitions
have a higher value of $N(X)/N_H f$ than does \nad 5892, 5898. 
Second, expected ionization corrections favor singly-ionized 
Mg, Fe, and Na over their neutral species.
Third, we find a larger cloud covering fraction
for the singly-ionized lines than for neutral lines.

We emphasize that the highest velocity absorption detected in 
\mgII appears to be limited by covering fraction. At continuum
$SNR \sim 10$, the absorption trough blends with the continuum
where $C_f(v) \sles 0.1$, regardless of column density. The
gas at the largest radii may well go undetected by typical 
absorption-line measurements.  For example, if we assume unity covering 
fraction at $R_0$, then gas beyond $R \sim 3.2 R_0$ cannot be
detected if $C_f$ falls as $R^{-2}$. In the more realistic case, 
$C_f \propto R^{-2/3}$, the covering fraction drops to 0.1 at $32 R_0$.
For a launch radius $R_0 \approx 200$~pc, our measurement would
detect gas out ot 6.4~kpc. This distance exceeds the extent of the
\Ha filaments in many nearby starburst galaxies but remains
considerably smaller than the recently detected soft, X-ray halos
in ULIRGs (Sciortino \& Martin, in prep.).  Spectroscopy of
background light sources at projected separations of just a few
tens of kpc from starburst and post-starburst galaxies should be 
more effective than line-of-sight studies for determining the 
spatial extent of the low-ionization outflow, but such studies
need to account for the geometrical dilution of the clouds.

\section{SUMMARY AND CONCLUSIONS} \label{sec:summary}

We presented the first comparison of optical and near-ultraviolet absorption
troughs in ULIRG spectra. We detected outflows in 5 ULIRGs in ground-state absorption
from \nad, \mgI, \mgII, and \feII. Previous observations of ULIRG outflows have been 
limited to the \nad\ doublet, and blending of the  \nad 5890, 5896 lines complicated 
measurement of the absorption trough shape. We summarize the primary, empirical results.

\begin{itemize}

\item
Comparison of the unblended, doublet components of \mgII and \feII distinguish the effects of 
optical depth and covering fraction in determining the shape of the absorption troughs. The
high optical depth in these transitions {\it at all outflow velocities} places a lower limit
on the column density. The non-zero intensity  requires partial coverage of the continuum source 
by the low-ionization gas over a broad velocity range.

\item
The covering fraction in all four ions decreases as the outflow velocity increases beyond the 
velocity of minimum intensity, or equivalently the velocity of maximum covering fraction. At a 
given velocity, the \mgI covering fraction is roughly half that measured in \mgII (and \feII),
and the covering fraction in \nad is less than or equal to that measured in \mgI.

\item
Accounting for these differences in ionic covering fraction, and taking spectral SNR into
consideration, we detect \mgII, \feII, \mgI, and \nad absorption over the same velocity range.
The decrease in covering fraction with increasing velocity suppresses the absorption signature
of the highest velocity gas. The higher covering fraction of \mgII (and \feII) yields 
detection to higher velocity than indicated by \mgI (or \nad). Any comparison of terminal
velocities measured from \nad and \mgII must take the bias introduced by the velocity-dependent 
covering fraction into consideration.  Many measurements exist for  \nad at $z \sim 0.6$,
whereas ground-based, optical spectrographs can measure outflows with the \mgII doublet over 
the broad redshift range from $\sim 0.25$ to $z \sim 2.5$. Caution should be exercised 
when examining evolution in outflow properties between these samples.  For completeness,
we point out that our spectra do not cover high-ionization transitions like \oVI 1032, 1038,
which reveal higher-velocity gas in some starburst outflows (Grimes \et 2009).

\item
The absence of \feIIe lines, principally the $\lambda 2612$ transition, place an upper limit 
on the volume density of $n_e < 10^{3.5}$\cm3, and likely $ < 10^{3.1}$\cm3. When present in quasar 
outflows, low-ionization absorption troughs from excited or metastable states indicate much 
higher volume density $n_e = 10^{4.4}$\cm3 with less than 20\% scatter (Korista \et 2008; Arav \et 2008). 
In the lower density ULIRG outflows, \mgII will be the dominant ionization state in the low-ionization 
outflow over a broad range of spectral hardness. In contrast, at these densities the \nad ionization fraction 
remains very sensitive to spectral hardness. We suggest that the harder radiation field in 
dwarf galaxies likely explains the lower fraction of dwarf starburst outflows detected in \nad relative
to ULIRGs, which are almost always detected in \nad (Martin 2005; Rupke \et 2005b).

\item 
We found  \mgII emission in two Sey~2 ULIRGs and He~I emission in all spectra.

\end{itemize}

These results provide new insight into the relationship of the low-ionization
outflow and the hot wind. We defer physically-motivated models of the absorption trough
shape to another paper (Martin 2009, in preparation)) 
but summarize key aspects of the emerging, physical picture here.

\begin{itemize}
\item
We associate the velocity of maximum covering fraction with that of a swept-up shell of interstellar gas 
at the time of blowout.  Factors motivating this interpretation include the Doppler shift of the
trough minimum (0 to -400\kms), the large width of the absorption troughs (up to 800\kms), the column 
density lower limits, and results from recent numerical simulations (Fujita \et 2009). Regardless 
of their physical origin, however, the large velocity width of the absorption troughs require 
contributions from multiple structures along the sightline. The relative shapes of the doublet 
troughs require these clouds or filaments to transition sharply (spatially) from opaque to 
optically thin gas. The neutral alkalai metals reside in the same kinematic structures as the 
singly ionized metals but fill a smaller fraction of that volume.

\item
The most significant result of our study may be the discovery
of a velocity-dependent covering fraction in low-ionization
outflows. The simplest interpretation is geometrical. The dilution associated 
with the spherical expansion of a population of absorbers causes their covering
fraction to decrease with increasing radius. We showed that the adiabatic
expansion of clouds in pressure equilibrium with the hot wind is not fast enough to offest this dilution.
{\it In the context of this physical scenario, our result implies that the  
high-velocity gas detected in the absorption trough is at larger radii 
than the lower velocity (and higher covering fraction) gas.} This 
mapping between velocity and relative radius indicates allows requires acceleration of
the low-ionization gas.  Either shell blowout or subsequent momentum-driving
by a hot wind and/or radiation pressure could cause this acceleration.

\item
In an alternative scenario, the shell fragments into clouds with a wide range of column 
densities. Due to momentum conservation, the ram pressure of the hot wind (or radiation pressure) 
would accelerate the lowest column density fragments to the highest terminal velocities. 
This relationship might even arise from blowout alone without requiring a momentum-driven wind phase.
Either way, the new empirical constraint, requiring 
lower covering fraction at higher velocity, would imply the covering fraction of low-ionization gas 
increases with increasing column density. We find the first, geometrical explanation more
appealing due to its simplicity.

\item
The lower limits on \nad and \feII column density provide the highest (i.e. strongest) lower
limits on the mass outflow rate in low-ionization gas. Better limits can be obtained by either
determing the \nad ionization fraction or observing bluer, \feII transitions with lower oscillator
strengths.

\item
Assuming that photoionization equilibrium likely sets the temperature of the low-ionization outflow,
$T_c \sim 10^4$~K, the upper limit on the volume density allows the pressure of the cold outflow to 
be as high as $P_c / k \sles 10^7$~K\cm3. Tighter constraints on the volume density would be valuable, as
very low density would be incompatible with pressure equilibrium with a hot wind, $P_c \approx P_h$, 
and favor a purely radiatively-driven outflow.

\end{itemize}

Our conclusion about the acceleration of the low-ionization outflow, while illustrative, is subject to an 
assumption about outflow geometry and a bias towards a shell origin for the low-ionization gas. Our empirical 
results, however, clearly challenge plausible dynamical models. Acceptable models need to reproduce the 
velocity-dependence of the low-ionization covering fraction, explain the lower covering fraction of  
neutral, alkalai metals relative to low-ionization species, and be consistent with the
observed constaints on both column density and volume density. With sufficient spectral sensitivity, all of
these outflow properties can be measured over a very broad redshift range, so we expect any evolutionary
effects with galaxy mass and/or cosmic time to eventually be measured.

\acknowledgements{We thank Nahum Arav, Tim Heckman, and Peng Oh for useful discussions
regarding the interpretation of these data. We thank Doug Edmonds for making his number density diagnostics
available to us and an anonyous referee for insightful comments. 
Special thanks to Amiel Sternberg
for his assistance in setting up STARS with the UVBlue stellar library.
This work was supported by the National Science 
Foundation under contract 080816, the David and Lucile Packard Foundation,  and the Aspen
Center for Physics.
}

\references

Arav, N. \et 1999a, \apj, 516, 27   

Arav, N. \et 2008, \apj, 681, 954 

Arav, N. \et 2005, \apj, 620, 665

Arav, N. \et 2001, \apj, 546, 140  

Borne, K. D. \et 2000, \apj, 529, L77 

Breitschwerdt, D. 2008, Nature, 452, 826 

Cooper, J. L. \et 2008, \apj, 674, 157

Dahlem, M., Weaver, K. A., \& Heckman, T. A. 1998, \apj, 118, 401 

Davies, R. I. \et 2007, \apj, 671, 1388 

DeYoung, D. S. \& Heckman, T. M. 1994, \apj, 431, 598

de Kool, M. \et 2002, \apj, 580, 54

Everett, J. \et 2008, \apj, 674, 258 

Fujita, A. \et 2009, \apj, 698, 693

Gabel, J. R. \et 2003 \apj, 583, 178

Grimes, J. P. \et 2009, \apjs, 181, 272 

Hanuschik, R. W. 2003, \aa, 407, 1157 

Heckman, T. \et 1990, \apjs, 74, 833

Heckman, T., Lehnert, M. D., Strickland D. K., \& Lee, A. 2000, \apjs, 129, 493 

Hoopes, C. G. \et 2005, \apj, 619, L99

Kennicutt, R. C. 1989 ApJ, 344, 685 [K89]

Kewley, L. J. \et 2006, \mn, 372, 961

Kim, D. C. \et 1998b \apj, 508, 627 

Kim, D.-C. \& Sanders, D. B. 1998, \apjs, 119, 41 [KS98] 

Kim, B.-C. \et 2002, \apjs, 143, 277 

Korista, K. T. \et 2008, \apj, 688, 108 

Leitherer, C. 1999, \apjs, 123, 3

Le Floc'h, E. \et 2005, \apj, 632, 169

Mac Low, M.-M., McCray, R., \& Norman, M. L. 1989, \apj, 337, 141

Marcolini, A. \et 2005, \mn, 362, 626

Martin, C. L. 1998, \apj, 506, 222

Martin, C. L. 1999, \apj, 513, 156

Martin, C. L. 2005, \apj, 621, 227 

Martin, C. L. 2006, \apj, 647, 222 

Martin, C. L., Kobulnicky, H. A., Heckman, T. M. 2002, \apj, 574, 663

McCarthy, J. K., Cohen, J. G., Butcher, B., Cromer, J., Croner, E., Douglas, W. R., Goeden, R. M., Grewal, T., Lu, B., Petrie, H. L., Weng, T., Weber, B., Koch, D. G., \& Rodgers, J. M. 1998 SPIE, 3355, 81

McKee, C. F. \& Ostriker, J. P. 1988, RvMP, 60, 1 

Morton, D. C.,  1991, \apjs, 77, 119

Morton, D. C.,  2003, \apjs, 149, 205

Murray, N.  Martin, C. L, Quataert, E \&  Thompson, T. A. 2007, \apj, 660, 211

Murray, N., Quataert, E., \& Thompson, T. A. 2005, \apj, 618, 569 

Oke, J. B., Cohen, J. G., Carr, M., Cromer, J., Dingizian, A. \& Harris, F. H. 1995, PASP, 107, 375

Oppenheimer \& Dav\'e 2006, \apj, 373, 1265 

Rodr\'iguez-Merino, L. H., Chavez, M., Bertone, E.,
\& Buzzoni, A. 2005, \apj, 626, 411  

Rupke, D. S., Veilleux, S., \& Sanders, D. B. 2002, \apj, 570, 588

Rupke, D. S., Veilleux, S., \& Sanders, D. B. 2005a, \apjs, 160, 87

Rupke, D. S., Veilleux, S., \& Sanders, D. B. 2005b, \apjs, 160, 115

Rupke D., Veilleux, S., \& Sanders, D. B. 2005d, \apj, 632, 751  

Sanders, D. B. \et 1988, \apj, 328, 35 

Sanders, D. B. \& Mirabel, F. 1996 \araa 34, 725

Savage, B. D. \& Sembach, K. R. 1996, \araa, 34, 279

Sato, T. \et 2009, \apj, 696, 214

Schwartz, C. S. \& Martin, C. L. 2004, \apj, 610, 20

Scott, J. E. 2004, \apjs, 152, 1  %

Shapley, A. E. \et 2003, \apj, 588, 65 

Shopbell, P. L. \& Bland-Hawthorne, J. 1998, \apj, 493, 129

Shull, M. 1993, in ASP Conf. Proceedings 35,Massive Stars: Their Lives in the Interstellar Medium, ed. J. P. Cassinelli \& E. B. Churchwell (San Francisco: ASP), 327

Shull, J. M. \& Van Steenberg, M. 1982, \apjs, 48, 95 

Socrates, A., Davis, S. W., \& Ramirez-Ruiz, E. 2008, \apj, 687, 202 

Somerville, R. S. \& Primack, J. R. 1999, \mn, 310, 1087 

Sternberg, A. 1998, \apj, 506, 721 

Sternberg, A., Hoffmann, T. L. \& Pauldrach, A. W. A.  2003, \apj, 599, 1333 

Strickland, D. K. \et 2002, \apj, 568, 689

Tenorio-Tagle, G., \& Bodenheimer, P. 1988, \araa, 26, 145 

Thornley, M. D. \et 2000, \apj, 539, 641 

Veilleux, S. \et  1999,  \apj 522, 113 

Veilleux, S. \et 2002, \apjs, 143, 315

Veilleux, S. \et 2006, \apjs, 643, 707

Veilleux, S. \et 2009a, \apjs, 182, 628 

Veilleux, S. \et 2009b, \apj, 700, 149  

Walter, F., Weiss, A., \& Scoville, N. 2002, \apj, 580, 21 

Weaver, R. \et 1977, \apj, 218, 377

Weiner, B. J. \et 2009, \apj, 692, 187


\setcounter{figure}{0}
\begin{figure*}[h]
\centering
\subfigure[ LRIS spectra of FSC0039-13]{
  \includegraphics[scale=0.7,angle=-90,clip=true]{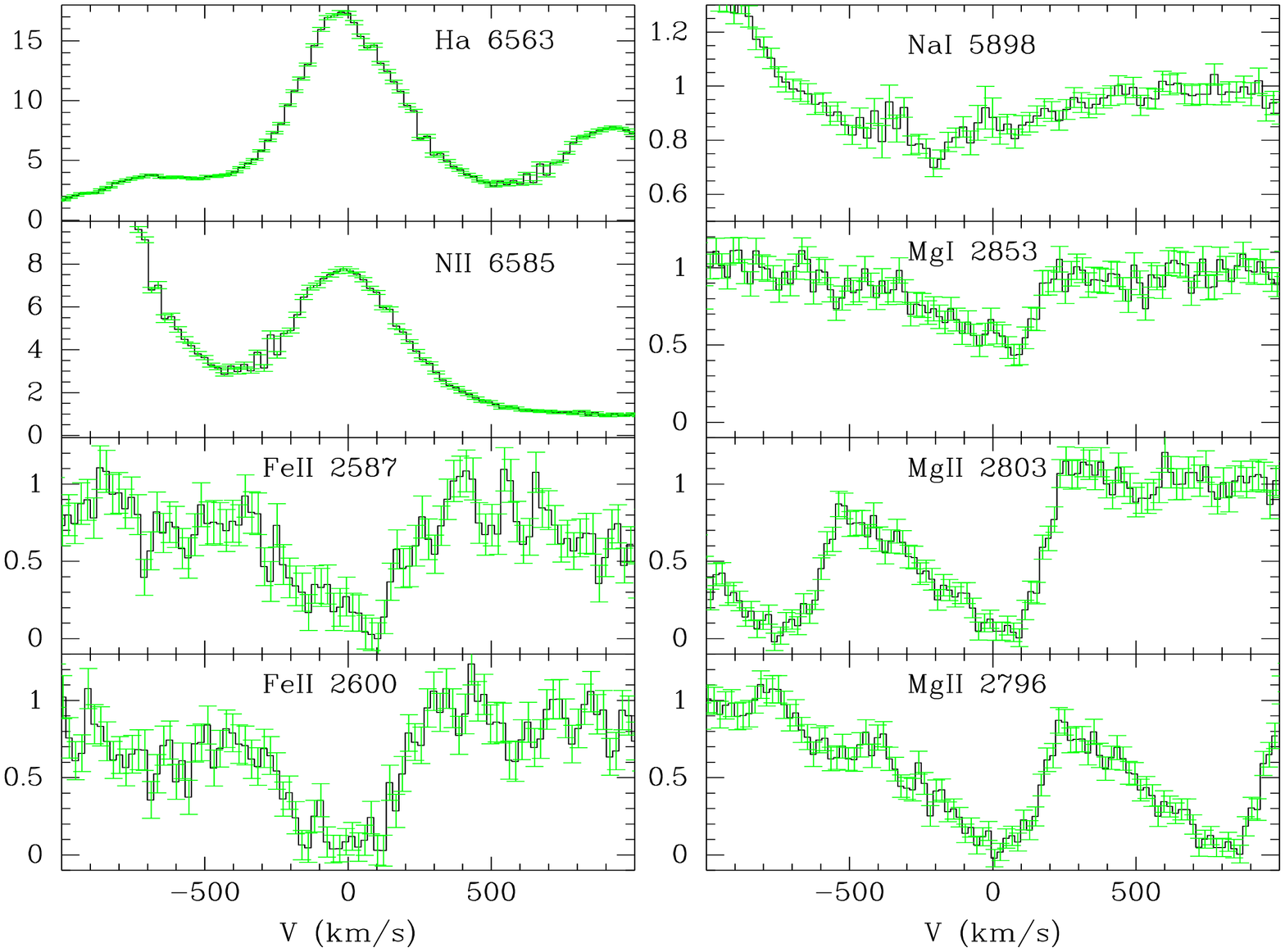}
}
\caption{Normalized intensity vs. velocity, where the velocity is relative 
to the systemic velocity determined from emission lines.
\label{fig:spec}
}
\end{figure*}

\addtocounter{figure}{-1}
\begin{figure*}[h]   
\addtocounter{subfigure}{1}
\subfigure[LRIS spectra of FSC1009+47]{
   \includegraphics[scale=0.7,angle=-90,clip=true]{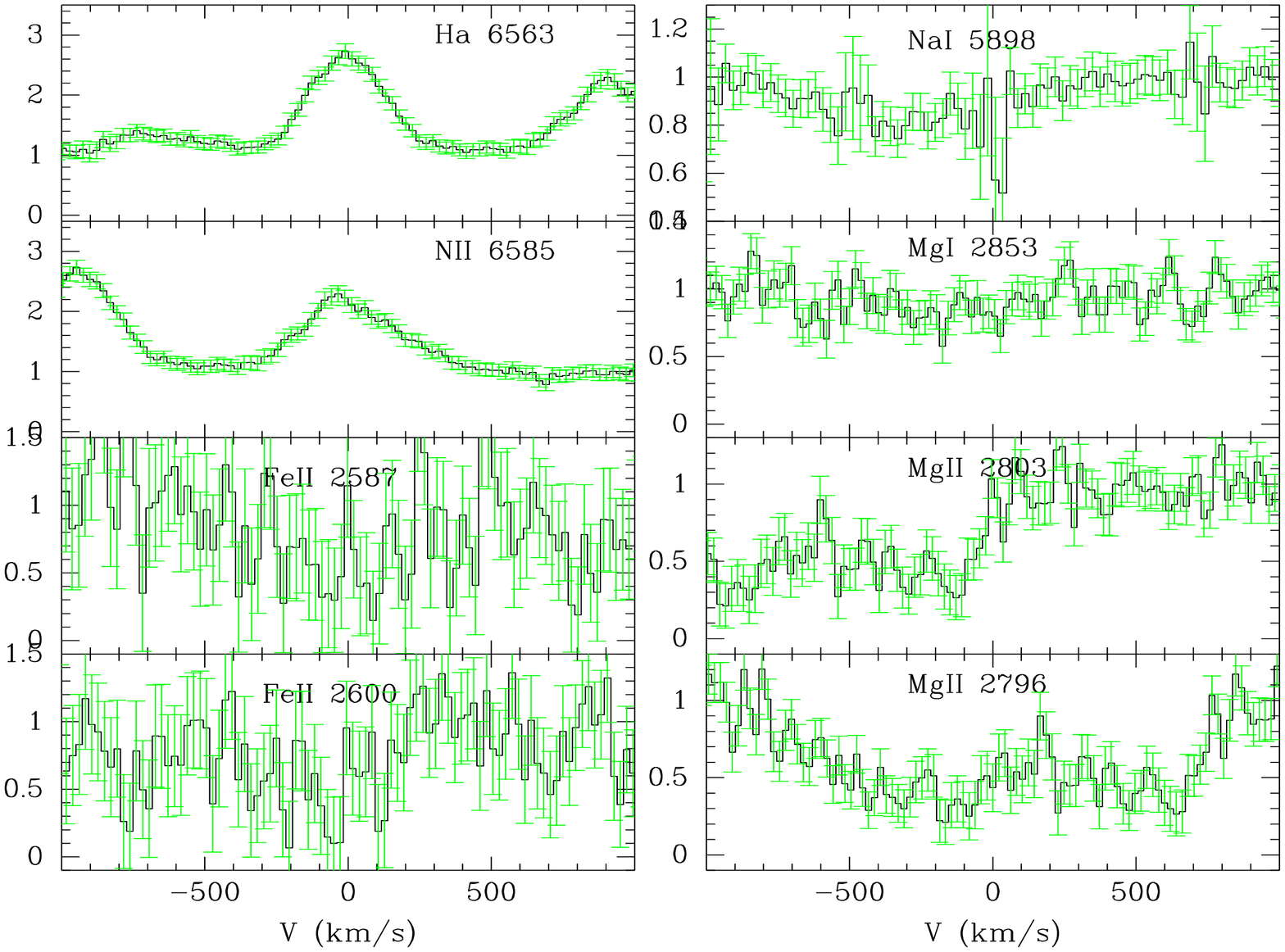}
  }
\caption{continued}
\end{figure*}

\addtocounter{figure}{-1}
\begin{figure*}[h]   
\addtocounter{subfigure}{1}
\subfigure[LRIS spectra of FSC1407+05]{
   \includegraphics[scale=0.7,angle=-90,clip=true]{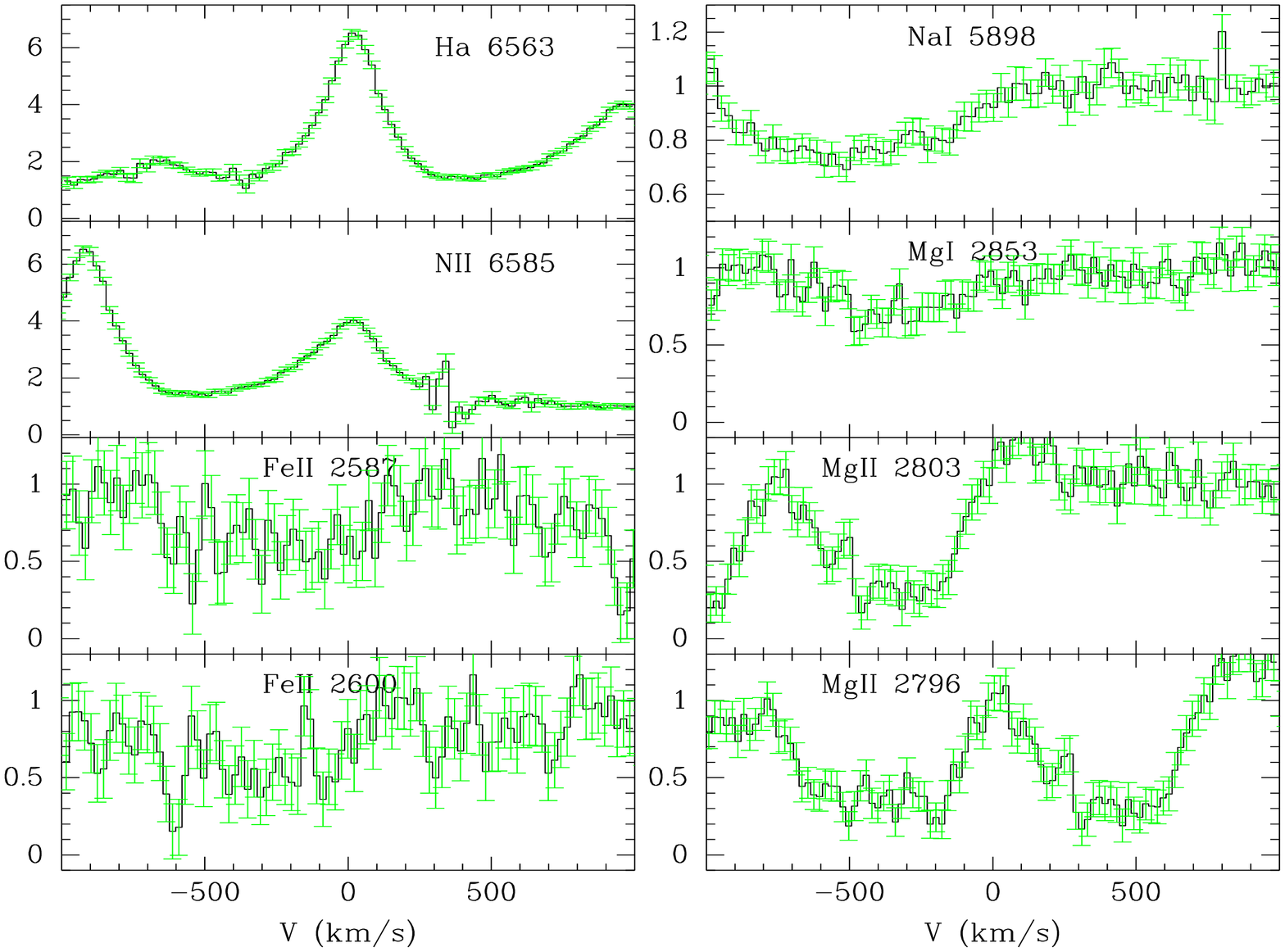}
  }
  \caption{continued}

\end{figure*}

\addtocounter{figure}{-1}
\begin{figure*}[h]  
\addtocounter{subfigure}{1}
\subfigure[LRIS spectra of FSC1630+15]{
   \includegraphics[scale=0.7,angle=-90,clip=true]{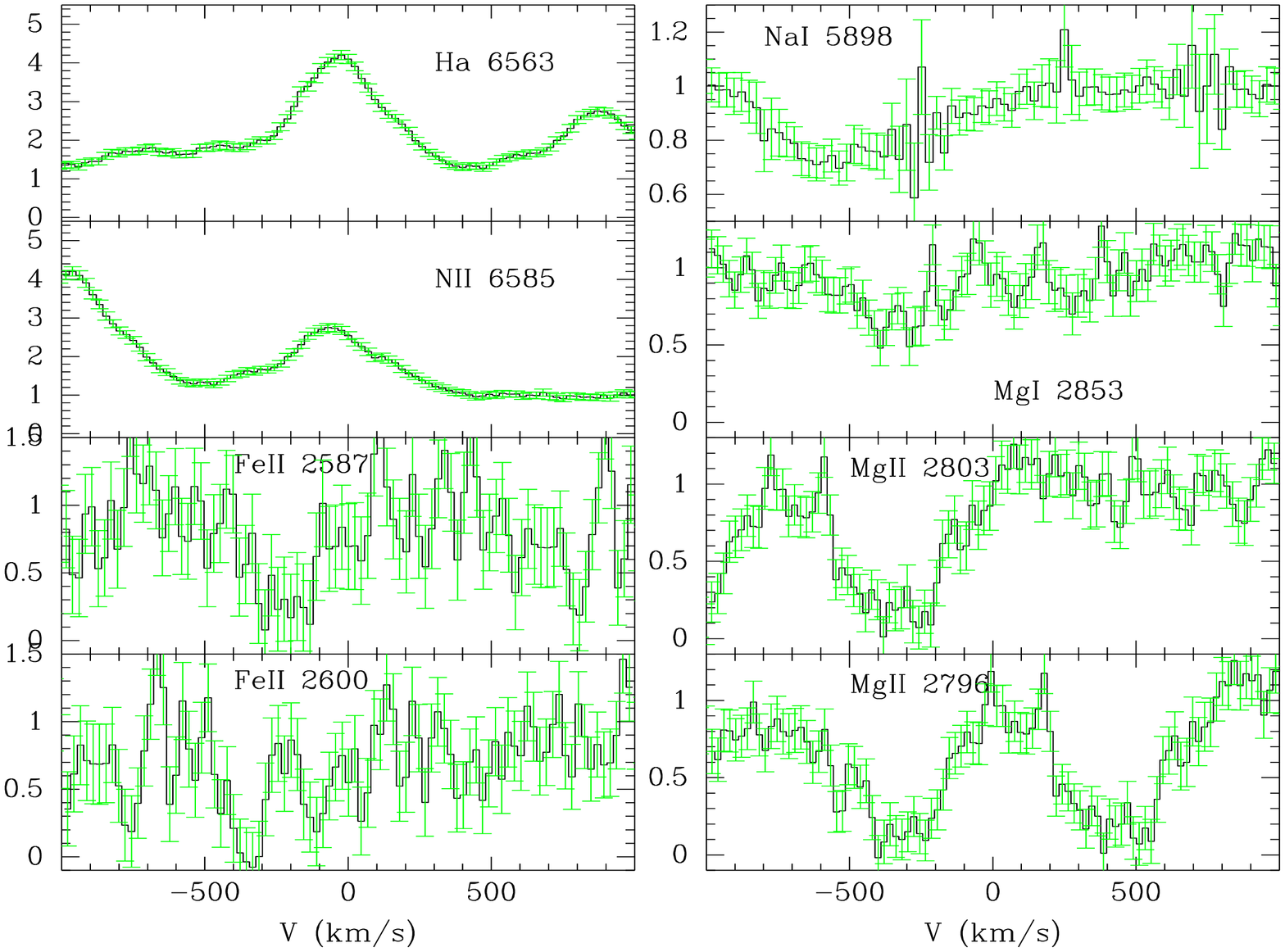}
  }
  \caption{continued}

\end{figure*}

\addtocounter{figure}{-1}
\begin{figure*}[h]
\addtocounter{subfigure}{1}
\subfigure[LRIS spectra of FSC2349+24]{
   \includegraphics[scale=0.7,angle=-90,clip=true]{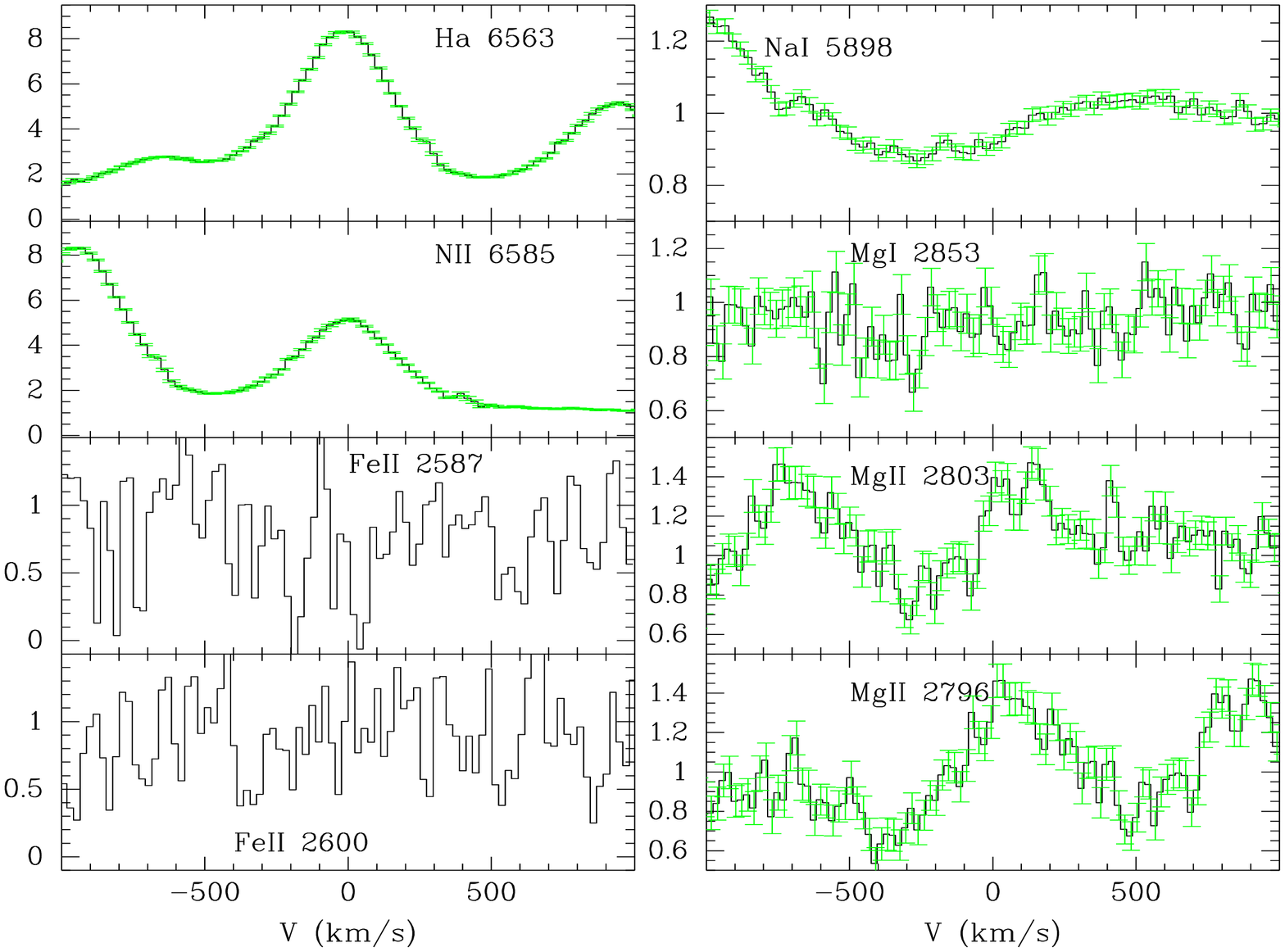}
  }
  \caption{continued}

\end{figure*}

\clearpage

\begin{deluxetable}{lllllll}
\tablewidth{0pt}
\caption{Properties of Galaxies}
\tablehead{
\colhead{Object} &
\colhead{f25/f60} &
\colhead{f60/f100} &
\colhead{Spectral Type} &
\colhead{$ \log (L_{IR}/ \lsun)$} &
\colhead{SFR-S} &
\colhead{SFR-C}
\\
\colhead{} &
\colhead{} &
\colhead{} &
\colhead{} &
\colhead{} &
\colhead{(\msunyr)} &
\colhead{(\msunyr)} 
\\
\colhead{(1)} &
\colhead{(2)} &
\colhead{(3)} &
\colhead{(4)} &
\colhead{(5)} &
\colhead{(6)} &
\colhead{(7)} 
}
\startdata
FSC00397-1312     & 0.180 & 0.963 & HII (V99)   & 12.81    & 1110    & 653 \\
FSC10091+4704  & 0.068 & 0.761 & LINER(V99,K98b) & 12.57    & 641     & 377 \\
FSC14070+0525  & 0.131 & 0.797 & Sey2 (K98b) & 12.66    & 789     & 464 \\
FSC16300+1558   & 0.047 & 0.744 & LINER (K98b)& 12.53    & 585     & 344 \\
FSC23498+2423  & 0.118 & 0.703 & Sey2 (V99)  & 12.31    & 352     & 207 \\ 
\hline	           	       	       		
\enddata
\tablecomments{
(1) ULIRG name.
(2) f25/f60 color computed from far-infrared fluxes in KS98.
(3)  f60/f100 color computed from far-infrared fluxes in [KS98].
(4) Spectral type from Kim \et (1998b) -- FSC1009+47, FSC1407+05, FSC1630+15,
-- and Veilleux \et (1999) -- FSC0039-13, FSC1009+47, FSC2349+24.
(5) $L_{IR}$ calculated using the prescription of Sanders \& Mirabel
(1996). For FSC objects, the luminosities from Kim \et (2002) and  Veilleux \et 
(1999), which assume luminosity distances based on $H_0 = 75$\kms ~Mpc$^{-1}$ 
and $q_0 = 0$, were converted to the cosmology used throughout this paper. 
(6) For Salpeter IMF  from 0.1 to 100\msun, the star formation rate is 
$SFR = L_{IR} / 5.8 \times 10^9 \lsun $, where $L_{IR}$ is
the bolometric luminosity (Kennicutt 1989).
(7) SFR for Chabrier IMF, i.e. K89 SFR by 1.7.
}
\label{tab:sample}
\end{deluxetable}


\begin{deluxetable}{lllllll}
\tablewidth{0pt}
\caption{Summary of Spectral Observations}
\tablehead{
\colhead{Object} &
\colhead{Redshift} &
\colhead{PA} &
\colhead{Res.} &
\colhead{SNR} &
\colhead{Date Observed} &
\colhead{Notes}
\\
\colhead{} &
\colhead{} &
\colhead{(deg)} &
\colhead{(km/s)} &
\colhead{} &
\colhead{} &
\colhead{} 
\\
\colhead{(1)} &
\colhead{(2)} &
\colhead{(3)} &
\colhead{(4)} &
\colhead{(5)} &
\colhead{(6)} &
\colhead{(7)} 
}
\startdata
FSC00397-1312 & 0.26171 &-54.0& 160   & 9.4   & 2007 Nov. 1 & Advanced Merger \\ 
FSC10091+4704 & 0.24508 &168.0& 130   & 5.0   & 2004 Jan. 26, Mar. 16-17 & Diffuse Merger \\ 
FSC14070+0525 & 0.26602 &-25.0& 110   & 8.2   & 2004 Jan. 26, Mar. 16-17 & Advanced Merger \\ 
FSC16300+1558 & 0.24200 &-45.0& 155   & 6.2   & 2004 Mar. 16-17, 2007 Oct. 6 & Pre-Merger, Separation 4.4~kpc  \\ 
FSC23498+2423 & 0.21249 &-45.9& 160   & 6.4   & 2007 Nov. 1 & Pre-Merger, Separation 14~kpc \\ 
\hline					   
\enddata
\tablecomments{
(1) Target. 
(2) Redshift measured from emission lines including
\Ha, [NII] 6584,48, [SII] 6717,31, and [OI] 6300,64. 
The absolute velocities have not been corrected
to the Local Standard of Rest. 
(3) Position angle of the longslit measured east of north.
(4) Spectral resolution defined by the measured
full width at half maximum intensity of the
arc lamp lines. If frames from different runs are
combined, then the poorer of the two resolutions is listed.
(5) SNR per pixel is measured in a 100\AA\ bandpass immediately
blueward of MgII. 
(6) Date(s) observed.
(7) The Veilleux \et (2002) merger classification is compared to
the structure observed along the longslit.
}
\label{tab:data} 
 \end{deluxetable}

\clearpage

\begin{turnpage}
\begin{deluxetable}{llllllllllllll}
\tabletypesize{\scriptsize}
\tablecaption{Fitted Covering Fractions}
\tablewidth{0pt}
\tablehead{
\colhead{FSC} &
\colhead{V(km/s)} &
\colhead{$C_f(\mgII )$} &
\colhead{$\tau_0(2803)$} &
\colhead{$\chi^2_{\nu}(\mgII)$} &
\colhead{$C_f(\mgI )$} &
\colhead{$\tau_0(2853)$} &
\colhead{$\chi^2_{\nu}(mgI)$}  &
\colhead{$C_f(\nad )$} &
\colhead{$\tau_0(5898)$} &
\colhead{$\chi^2_{\nu}(\nad)$} &
\colhead{$C_f(\feII )$}  &
\colhead{$\tau_0(2587)$} &
\colhead{$\chi^2_{\nu}(\feII)$} 
\\
\colhead{(1)} &
\colhead{(2)} &
\colhead{(3)} &
\colhead{(4)} &
\colhead{(5)} &
\colhead{(6)} &
\colhead{(7)} &
\colhead{(8)} &
\colhead{(9)} &
\colhead{(10)} &
\colhead{(11)} &
\colhead{(12)} &
\colhead{(13)} &
\colhead{(14)}  
}
\startdata
0039-13   &    83  & $0.74 \pm 0.03$  &    17.7 &    1.136   &  $0.41 \pm  0.22$   &	 4     & 1.417   & $  0.17 \pm 0.02 $   &  4    &   2.335 &   $  0.99 \pm 0.04 $      &  2.8     &  2.452    \\
0039-13   &   -2.6 & $0.83 \pm 0.03$  &    17.7 &    1.136   &  $0.57 \pm  0.38$   &	4     & 1.417    & $  0.04 \pm 0.02 $   &  4    &   2.335 &   $  0.90 \pm 0.05 $      &  2.8     &   2.452   \\
0039-13   &   -135 & $0.47 \pm 0.06$  &    17.7 &    1.136   &  $0.20 \pm  0.08$   &	4     & 1.417    & $  0.11 \pm 0.02 $   &  4   &    2.335 &   $  0.72 \pm 0.05 $       &  2.8     &   2.452  \\
0039-13   &   -209 & $0.45 \pm 0.05$  &    17.7 &    1.136   &  $0.24 \pm  0.07$   &	4     & 1.417    & $  0.05 \pm 0.02 $   &  4   &    2.335 &   $  0.50 \pm 0.06 $       &  2.8     &   2.452  \\
0039-13   &  -379  & $0.33 \pm 0.02$  &   17.7	&   1.136    &  $0.19 \pm  0.04$   &	4     & 1.417    & $  0.09 \pm 0.01 $   &   4   &   2.335 &	 $  0.31 \pm 0.04 $      &  2.8     &  2.452    \\
0039-13   &   -571 & $0.27 \pm 0.02$  &    17.7 &    1.136   &  $0.21 \pm  0.03$   &	4     & 1.417    & $  0.07 \pm 0.01 $   &  4    &   2.335 &	 $  0.43 \pm 0.04 $       &  2.8     &  2.452   \\
\hline  	    		     		       	      		      		            	      		       		   	                               
1009+47   &   -104 & $0.66 \pm 0.05$    &  5	&    0.955   &  $0.17 \pm  0.11$    &  3.4      &  1.243 & $  0.16 \pm 0.05 $    &   3 &   0.372  &  \nodata          & \nodata     & \nodata            \\        
1009+47   &   -361 & $0.55 \pm 0.05$    &  5	&    0.955   &  $0.11 \pm  0.08$    &  3.4      &  1.243 & $  0.08 \pm 0.05 $     &   3 &  0.372  &  \nodata          & \nodata     & \nodata            \\        
1009+47   &   -239 & $0.51 \pm 0.05$    &  5	&    0.955   &  $0.25 \pm  0.10$     &  3.4     &  1.243 & $  0.17 \pm 0.04 $     &   3 &  0.372  &  \nodata          & \nodata     & \nodata         \\        
1009+47   &   -507 & $0.51 \pm 0.04$    &  5	&    0.955   &  $0.07 \pm  0.07$     &  3.4     &  1.243 & $  0.02 \pm 0.04 $     &   3 &  0.372  &  \nodata          & \nodata     & \nodata         \\        
1009+47   &   -692 & $0.38 \pm 0.04$    &  5	&    0.955   &  $0.10 \pm  0.06$    &  3.4      &  1.243 & $  0.04 \pm 0.06 $     &   3 &  0.372  &  \nodata          & \nodata     & \nodata         \\        
\hline  	    		     		       	      		      		            	      		       		   	                                    
1407+05   &    -89 & $0.29 \pm 0.04$    &  10	&    0.989   &   $0.16 \pm  0.08$    &  6.8     &  1.045 & $  0.11 \pm 0.02 $      &   3&  0.702  &  $ 0.52 \pm  0.07$  & 1.6 &   1.232               \\
1407+05   &   -187 & $0.40 \pm 0.07$    &  10	&    0.989   &   $0.08 \pm  0.07$    &  6.8     &  1.045 & $  0.10 \pm 0.02 $      &   3&  0.702  &  $ 0.14 \pm  0.11$  & 1.6 &   1.232               \\
1407+05   &   -265 & $0.62 \pm 0.04$    &  10	&    0.989   &   $0.28 \pm  0.09$     &  6.8    &  1.045 & $  0.17 \pm 0.02 $     &   3&  0.702	  &  $ 0.55 \pm  0.08$  & 1.6 &   1.232                  \\
1407+05   &   -434 & $0.67 \pm 0.04$    &  10	&    0.989   &   $0.37 \pm  0.05$    &  6.8     &  1.045 & $  0.20 \pm 0.01 $      &   3& 0.702   &  $ 0.49 \pm  0.07$  & 1.6 &   1.232                   \\
1407+05   &   -615 & $0.51 \pm 0.03$    &  10	&    0.989   &   $0.15 \pm  0.04$    &  6.8     &  1.045 & $  0.14 \pm 0.02 $      &   3&  0.702  &  $ 0.59 \pm  0.07$  & 1.6 &   1.232                       \\
\hline  	    		     		       	      		      		            	                                                                
1630+15   &   -114 & $0.28 \pm 0.05$    &   5	&    1.086   &  $0.10 \pm  0.09$      &  3.4    &   1.014& $  0.13 \pm 0.03 $    &   3 &  0.491   & \nodata     & \nodata   &  \nodata                     \\
1630+15   &   -252 & $0.81 \pm 0.05$    &   5	&    1.086   &  $0.24 \pm  0.12$      &  3.4    &   1.014& $  0.21 \pm 0.03 $    &   3&    0.491  & \nodata     & \nodata    & \nodata                     \\
1630+15   &   -362 & $0.82 \pm 0.05$    &   5	&    1.086   &  $0.42 \pm  0.14$      &  3.4    &   1.014& $  0.16 \pm 0.04 $    &   3&    0.491  & \nodata     & \nodata    & \nodata                     \\
1630+15   &   -491 & $0.60 \pm 0.05$    &   5	&    1.086   &  $0.22 \pm  0.07$     &  3.4     &   1.014& $  0.12 \pm 0.03 $    &    3&   0.491  & \nodata     & \nodata    & \nodata                     \\
1630+15   &   -676 & $0.21 \pm 0.04$    &   5	&    1.086   &  $0.08 \pm  0.05$     &  3.4     &   1.014& $  0.07 \pm 0.04 $    &    3&   0.491  & \nodata     & \nodata    & \nodata                     \\
\hline  	    		     		       	      		      		            	                                                                
2349+24   &    -71 & $0.27 \pm 0.03$    &   2	&    2.247   &  $0.12 \pm  0.07$     &  1.4     &   2.254&  $  0.08 \pm 0.007$    &   5&    4.259 & \nodata     & \nodata	& \nodata	                 \\
2349+24   &   -182 & $0.28 \pm 0.03$    &   2	&    2.247   &  $0.04 \pm  0.05$      &  1.4    &   2.254&  $  0.05 \pm 0.007$    &   5&    4.259 & \nodata     & \nodata      &  \nodata                       \\
2349+24   &   -298 & $0.38 \pm 0.03$    &   2	&    2.247   &  $0.30 \pm  0.07$    &  1.4      &   2.254&  $  0.02 \pm 0.007$    &   5&    4.259 & \nodata     & \nodata      &  \nodata                       \\
2349+24   &   -423 & $0.27 \pm 0.03$    &   2	&    2.247   &  $0.27 \pm  0.06$      &  1.4    &   2.254&  $  0.02 \pm 0.007$    &   5&    4.259 & \nodata     & \nodata      &  \nodata                       \\
2349+24   &   -591 & $0.17 \pm 0.03$    &   2	&    2.247   &  $0.17 \pm  0.05$      &  1.4    &   2.254&  $  0.008\pm 0.007$    &   5&    4.259 & \nodata     & \nodata      &  \nodata                       \\
\hline
\enddata
\tablecomments{(1) Galaxy. 
(2) Doppler shift of component in \kms. All components have a Doppler parameter $b \equiv 50$\kms, or $b_{50} \equiv 1$.
(3) Fitted covering fraction for \mgII
(4) Minimum optical depth at line center for \mgII 2803. The lower limit on the ionic column density is $N(\mgII) > 3.90 \times 10^{13} {\rm ~cm}^{-2} \tau_{0} b_{50}$.
(5) Fit statistic for the \mgII doublet.
(6) Fitted covering fraction for \mgI
(7) Minimum optical depth at line center for \mgI 2853. The lower limit on the ionic column density is $N(\mgI) > 6.39 \times 10^{12} {\rm ~cm}^{-2} \tau_{0} b_{50}$.
(8) Fit statistic for the \mgI line.
(9) Fitted covering fraction for \nad.
(10) Minimum optical depth at line center for \nad 5898. The lower limit on the ionic column density is $N(\nad) > 1.78 \times 10^{13} {\rm ~cm}^{-2} \tau_{0} b_{50}$.
(11) Fit statistic for the \nad doublet.
(12) Fitted covering fraction for \feII.
(13) Minimum optical depth at line center for \feII 2600. The lower limit on the ionic column density is $N(\feII) > 1.87 \times 10^{14} {\rm ~cm}^{-2} \tau_{0} b_{50}$.
(14) Fit statistic for the \feII doublet.
}
\label{tab:fit}

\end{deluxetable}
\end{turnpage}

\end{document}

%% file: mymac.tex
\newcommand{\nad}{\hbox{{\rm Na}\kern 0.1em{\sc i}~}}
\newcommand{\mgII}{\hbox{{\rm Mg}\kern 0.1em{\sc ii}~}}
\newcommand{\mgI}{\hbox{{\rm Mg}\kern 0.1em{\sc i}~}}
\newcommand\mnII{\hbox{{\rm Mn}\kern 0.1em{\sc ii}~}}
\newcommand{\feII}{\hbox{{\rm Fe}\kern 0.1em{\sc ii}~}}
\newcommand{\feIIe}{\hbox{{\rm Fe}\kern 0.1em{\sc ii}$^*$~}}
\newcommand{\feI}{\hbox{{\rm Fe}\kern 0.1em{\sc i}~}}
\newcommand{\oVI}{\hbox{{\rm O}\kern 0.1em{\sc vi}~}}

\def\H7{\mbox {$h_{0.7}$}}

\def\IZw18{I~Zw~18}
\def\m82{M82}

\def\h{\mbox {\rm H}}

\def\msun{\mbox {${\rm ~M_\odot}$}}

\def\lsun{\mbox {${~\rm L_\odot}$}}
\def\msunyr{\mbox {$~{\rm M_\odot}$~yr$^{-1}$}}

\def\Ha{\mbox {H$\alpha$~}}

\def\h0{\mbox {~H$_0$}}
\def\q0{\mbox {~q$_0$}}
\def\o3hb{[OIII]$\lambda5007$~/~H$\beta$~}
\def\O1ha{[OI]$\lambda6300$~/~H$\alpha$~}

\def\s2ha{[SII]$\lambda\lambda6717,31$~/~H$\alpha$~}
\def\2z2{HeII~$\lambda4686$~}
\def\z7{[NII]~$\lambda6583$ }
\def\N2{[NII]~$\lambda6583$~/~H$\alpha$~}
\def\16z2{[SII]~$\lambda\lambda6717, 6731$ }

\def\asec{\ifmmode {'' }\else $''~$\fi}  
\def\amin{\ifmmode {' }\else $'~$\fi}    
\def\arcsper{\ifmmode \rlap.{'' }\else $\rlap{.}'' $\fi} 
\def\arcmper{\ifmmode \rlap.{' }\else $\rlap{.}' $\fi} 
\def\sles{\lower2pt\hbox{$\buildrel {\scriptstyle <}
   \over {\scriptstyle\sim}$}} 
\def\sgreat{\lower2pt\hbox{$\buildrel {\scriptstyle >}
    \over {\scriptstyle\sim}$}} 
\def\kms{~km~s$^{-1}$~}

\def\cm3{~cm$^{-3}$}
\def\col{~cm$^{-2}$}
\def\mpc3{~Mpc$^{3}$}
\def\mpc-3{~Mpc$^{-3}$}

\def\um{~${\mu}$m~}
\def\fig{{Figure}}

\def\et{{\rm et\thinspace al.}\ }   

\def\apj{ApJ}
\def\apjs{ApJS}

\def\mn{MNRAS}

\def\aa{A\&A}